\documentclass[]{aastex631}
\usepackage{amsmath, amssymb}
\usepackage{booktabs}
\usepackage[para,online,flushleft]{threeparttable}
\usepackage{comment}

\newcommand{\be}{\begin{equation}}
\newcommand{\ee}{\end{equation}}
\newcommand{\diff}{\ensuremath{\mathrm{d}}}

\shorttitle{Multi-messenger puzzle of NGC 1068}
\shortauthors{Eichmann et al.}

\graphicspath{{./}{figures/}}

\begin{document}

\title{Solving the multi-messenger puzzle of the AGN-starburst composite galaxy NGC 1068}

\author{Bj\"orn Eichmann}
\affiliation{Norwegian University for Science and Technology (NTNU), Institutt for fysikk, Trondheim, Norway}
 \affiliation{Ruhr-Universit\"at Bochum, Theoretische Physik IV, Fakult\"at f\"ur Physik und Astronomie, Bochum, Germany}
\affiliation{Ruhr Astroparticle and Plasma Physics Center (RAPP Center), Bochum, Germany}

\author{Foteini Oikonomou}
\affiliation{Norwegian University for Science and Technology (NTNU), Institutt for fysikk, Trondheim, Norway}
  
\author{Silvia Salvatore}
 \affiliation{Ruhr-Universit\"at Bochum, Theoretische Physik IV, Fakult\"at f\"ur Physik und Astronomie, Bochum, Germany}
\affiliation{Ruhr Astroparticle and Plasma Physics Center (RAPP Center), Bochum, Germany}
  
\author{Ralf-Jürgen Dettmar}
  \affiliation{Ruhr-Universit\"at Bochum, Astronomical Institute, Fakult\"at f\"ur Physik und Astronomie, Bochum, Germany}
\affiliation{Ruhr Astroparticle and Plasma Physics Center (RAPP Center), Bochum, Germany}

\author{Julia Becker Tjus}
 \affiliation{Ruhr-Universit\"at Bochum, Theoretische Physik IV, Fakult\"at f\"ur Physik und Astronomie, Bochum, Germany}
\affiliation{Ruhr Astroparticle and Plasma Physics Center (RAPP Center), Bochum, Germany}

\begin{abstract}
Multi-wavelength observations indicate that some starburst galaxies show a dominant non-thermal contribution from their central region. These active galactic nuclei (AGN)-starburst composites are of special interest, as both phenomena on their own are potential sources of highly-energetic cosmic rays and associated gamma-ray and neutrino emission. In this work, a homogeneous, steady-state two-zone multi-messenger model of the non-thermal emission from the AGN corona as well as the circumnuclear starburst region is developed and subsequently applied to the case of NGC\,1068, which has recently shown some first indications of high-energy neutrino emission. Here, we show that the entire spectrum of multi-messenger data - from radio to gamma-rays including the neutrino constraint - can be described very well if both, starburst \emph{and} AGN corona, are taken into account. Using only a single emission region is not sufficient. 
\end{abstract}

\section{Introduction} \label{sec:intro}
The high rate of star formation and supernova explosions together with their relatively large abundance make starburst galaxies some of the most promising sources of high-energy cosmic rays (CRs) in the nearby Universe. 
Multiwavelength observations reveal that in some starburst galaxies the dominant non-thermal emission component originates in their central regions, indicating the presence of an active super-massive black hole. These active galactic nuclei (AGN)-starburst composites are of special interest, as both phenomena on their own are potential sources of highly-energetic CRs which may contribute to the extragalactic CR component observed at Earth, whose origin remains unknown. 

The \emph{Fermi}-Large Area Telescope (LAT) has detected high-energy gamma-rays from a number of nearby starburst galaxies~\citep{a5:fermi_starbursts2012}, whereas a handful of those (M82, NGC 253, NGC 1068) have been detected all the way up to TeV gamma-ray energies with imaging atmospheric Cherenkov telescopes~\citep{2009Sci...326.1080A,2009Natur.462..770V,HESS:2018yqa,MAGIC2019}, confirming that at least a fraction of the starburst galaxy population accelerate particles to very high energies. In addition, starburst galaxies have long been considered prime environments for high-energy neutrino production, e.g.~\citet{Loeb:2006tw,becker2009}, due to having higher magnetic fields and gas densities than Milky Way-like galaxies. A first observational hint of possible neutrino production in the starburst-AGN composite NGC\,1068 has been seen in 10 years of data from the IceCube Neutrino Observatory: an excess of neutrinos has been found in a region centered $0.35^{\circ}$ away from the coordinates of NGC\,1068 after a catalogue-based search using ten years of events from the point source analysis~\citep{a5:IceCube2020}. The excess is inconsistent with background expectations at the 2.9$\sigma$ level after accounting for trials. NGC\,1068 is the brightest and one of the closest Seyfert type 2 galaxies. Due to its proximity, at a distance of 14.4 Mpc~\citep{2004MNRAS.350.1195M}, it forms a prototype of its class. It has also been predicted as one of the brightest neutrino sources in the northern hemisphere~\citep{Murase:2016gly}. 

In order to understand the emission mechanisms active in these composite sources in general, and in NGC\,1068 in particular, and the possible connection to neutrinos, it is necessary to distinguish the non-thermal emission from the different acceleration sites. In particular in the case of NGC\,1068, there are strong observational hints for multiple emission sites: firstly, in the near-infrared to radio the ALMA experiment has observed a strong flux cutoff towards smaller frequencies emerging from the inner parsecs of that source \citep{ALMA2016,a5:ALMA2019a}. Further non-thermal radio emission is associated with the more extended region of a few hundreds of parsecs~\citep{WilsonUlvestad1982,Sajina+2011}. Secondly, the gamma-ray flux extends out to 100\,GeV~\citep{MAGIC2019}. 

 Indications that a one-zone model is not enough to explain the multi-messenger emission from NGC1068 have been present for several years. Using radio and gamma-ray observations, it has been shown in previous works, such as \cite{a5:eichmann2016, a5:Yoast-Hull-etal-2014}, that the circumnuclear starburst environment alone is not powerful enough (by a factor $\lesssim 10$) to account for the observed gamma-ray luminosity. Since NGC\,1068 also possesses a large-scale jet as indicated by centimeter radio observations \citep[e.g.][]{WilsonUlvestad1982,Gallimore+2004,Gallimore+2006} it has been suggested \citep{Lenain+2010} that the observed gamma-rays originate in these jets as a result of CR electrons inverse Compton scattering off the infrared (IR) radiation of the surrounding environment. Another possible origin of the observed gamma-ray emission that has been suggested by \cite{Lamastra+2019} are CR particles accelerated by the AGN-driven wind that is observed in the circumnuclear molecular disk of NGC 1068. This model predicts a rather hard spectrum extending up to a several TeV which, however, can be excluded due to the upper limits by the MAGIC telescope \citep{MAGIC2019}. 
The potential neutrino emission indicated from the IceCube excess at about 1\,TeV poses another problem, as it is at least an order of magnitude stronger than the GeV photon flux. These different components of the high energy phenomena can hardly be explained by a single zone model -- a high energy neutrino signal with a missing gamma-ray counterpart indicates an optically thick source environment such as the AGN corona \citep{a5:Inoue+2020, a5:Murase+2020, Kheirandish+2021}, whereas the presence of a gamma-ray flux up to some tens of GeV that is only slightly steeper than $\propto E^{-2}$ can only be explained by a second source environment.
In this work we model the multi-messenger emission of NGC\,1068 in the context of a two-zone model which considers the AGN corona and circumnuclear starburst region. In Section~\ref{sec:AGN-starburst-model} we present the details of the two-zone model and the formalism employed to model the multi-messenger emission of the source. In Section~\ref{sec:results} we present our fitting approach as well as its results and conclude in Section~\ref{sec:conclusions}. 

\section{The two-zone AGN-starburst model} \label{sec:AGN-starburst-model}
\begin{figure}[htbp]
    \centering
    \includegraphics[width=0.99\linewidth]{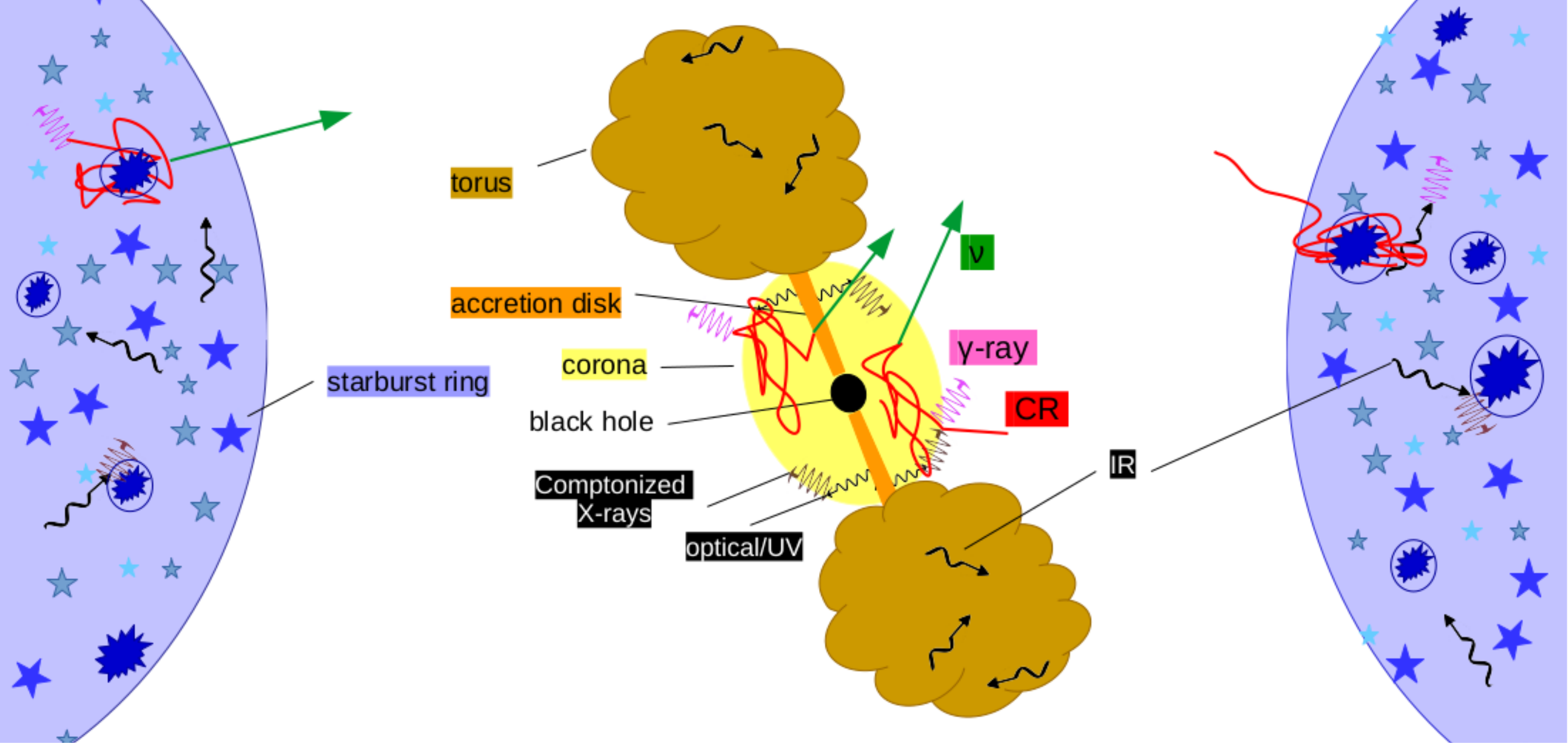}
    \caption{Sketch of the two-zone AGN-starburst model. The sizes of the regions are not to scale as the (inner) coronal region typically extends out to about some hundreds of $\mu\text{pc}$, whereas the starburst ring is located at about 1 kpc distance from the nucleus. } 
    \label{fig:AGNstarburst-sketch} 
\end{figure}

The multi-messenger observations of NGC\,1068 dictate the need for multiple emission zones. In this work, we will capture its inner $\mu \text{pc}$s referring to a spherically symmetric structure for the corona of the AGN as well as an outer starburst ring with a radius of $\sim$ 1 kpc (see Fig.~\ref{fig:AGNstarburst-sketch}). Note, that in between these two emission sites NGC\,1068 shows strong indications of a jet structure on scales of up to about $1\,\text{kpc}$ \citep[e.g.][]{WilsonUlvestad1982,Gallimore+2004,Gallimore+2006}, which however is not included in this work.
Due to mathematical convenience we treat both spatial regions as homogeneous. For particle acceleration processes that take place on considerably shorter timescales than the energy loss in these zones, we can disentangle these processes and only describe the steady-state transport of non-thermal, accelerated electrons and protons. Hereby, we suppose that in both zones some acceleration mechanism yields a differential source rate $q(T)$ of relativistic protons and primary electrons that can be described by a power-law distribution in momentum space up to a certain maximal kinetic energy $\hat{T}$, which depends on the competing energy loss timescales in these zones. In case of the starburst zone, we suppose that a certain fraction $f_{\rm SN}$ of the total energy of the supernova --- that release about $10^{51}\,\text{erg}$ and occur with an approximate rate\footnote{Supposing a supernova rate $\nu_{\rm SN}\simeq 0.02[SFR/(1M_\odot/\text{yr})]\text{yr}^{-1}$ (note that \cite{Condon1992} suggested a value of $0.04$ instead of $0.02$ for normal galaxies) where the star formation rate $SFR\simeq 17[L_{\rm IR}/(10^{11}L_\odot)]\,M_\odot\text{yr}^{-1}$.} \citep{Veilleux+2005} $\nu_{\rm SN}\simeq0.34\,[L_{\rm IR}/(10^{11}\,L_\odot)]\text{yr}^{-1}$ dependent on the IR luminosity $L_{\rm IR}$ --- gets accelerated into CRs according to diffusive shock acceleration \citep[DSA, e.g.][]{a5:Drury1983,a5:Protheroe1999} by individual supernova remnants \citep[e.g.][and references therein]{a5:Bell2014}. In general, many starburst galaxies--- NGC\,253 is a prominent example---show a galactic superwind \citep[e.g.][and references therin]{Veilleux+2005} as a result of the large number of core-collapse supernovae. These winds introduce another source of acceleration\footnote{Note that in these phenomena also stochastic diffuse acceleration may become relevant due to the presence of a turbulent plasma within the wind bubbles.} \citep[e.g.][]{a5:Anchordoqui+1999, a5:Romero+2018}, however, we are not aware of any observational indications of such a superwind in the starburst ring of NGC\,1068. 
For the AGN corona, we suppose that a fraction $f_{\rm inj}\ll 1$ of the mass accretion rate $\dot{M}=L_{\rm bol}/(\eta_{\rm rad}c^2)$, with a radiation efficiency of $\eta_{\rm rad}=0.1$ \citep{Kato+2008book}, goes into relativistic protons via stochastic diffuse acceleration \citep[SDA, e.g.][and references therein]{a5:LemoineMalkov2020}. 
For both zones, the non-thermal primary electrons are normalized by the non-thermal proton rates due to the requested quasi-neutral total charge number of the injection spectra of primary CRs above a characteristic kinetic energy of $\check{T}\simeq 10\,\text{keV}$ \citep{Schlickeiser2002_book,a5:eichmann2016, Merten+2017}. Note that this corresponds to a quasi-neutral acceleration site, however CR transport can subesequently remove CR electrons and protons in different amounts from the non-thermal energy regime, nevertheless their charge stays conserved.
Transforming the source rates from momentum space into kinetic energy $T$, we obtain 
\begin{equation}
    q_{\rm p}(T)\equiv \frac{\diff N}{\diff V\, \diff T\,\diff t} = \frac{q_{p,0}}{\sqrt{\check{T}^2+2\check{T}E_{p,0}}}\,\frac{T+E_{p,0}}{\sqrt{T^2+2TE_{p,0}}}\,\left( \frac{T^2+2TE_{p,0}}{\check{T}^2+2\check{T}E_{p,0}}\right)^{-s/2}\,\exp[-T/\hat{T}]\,,
\end{equation} 
for the injected non-thermal protons, and 
\begin{equation}
    q_{\rm e^\pm}(T) = \frac{q_{e^-,0}}{\sqrt{\check{T}^2+2\check{T}E_{e,0}}}\,\frac{T+E_{e,0}}{\sqrt{T^2+2TE_{e,0}}}\,\left( \frac{T^2+2TE_{e,0}}{\check{T}^2+2\check{T}E_{e,0}}\right)^{-s/2}\,\exp[-T/\hat{T}] + q^{\rm 2nd}_{\rm e^\pm}(T)\,,
\end{equation}
for the non-thermal electrons ($e^-$) and positrons ($e^+$). Here, the latter term $q^{\rm 2nd}_{\rm e^\pm}(T)$ introduces the source rate of secondary electrons and positrons that are generated by hadronic interaction processes, as discussed in the following. 

Thus, the steady-state behavior of the differential non-thermal electron and proton density $n(T)$ in the AGN corona and the starburst zone, respectively, can be approximated by
\begin{equation}
-\frac{\partial}{\partial T}\left(\frac{T\,n(T)}{\tau_{\mathsf{cool}}(T)}\right)=q(T)-\frac{n(T)}{\tau_{\mathsf{esc}}(T)}\,.
\label{eq:teq0}
\end{equation}
Here, $\tau_{\mathsf{cool}}$ refers to the total continuous energy-loss timescale, which in case of the relativistic electrons is given by the inverse of the sum of the synchrotron (syn), inverse Compton (IC), non-thermal Bremsstrahlung (brems), and Coulomb (C) loss rates, according to
\begin{equation}
    \tau_{\mathsf{cool}}^{\rm (e)}=[(\tau_{\mathsf{syn}}^{\mathsf{(e)}})^{-1}+\tau_{\mathsf{ic}}^{-1}+\tau_{\mathsf{brems}}^{-1}+(\tau_{\mathsf{C}}^{\mathsf{(e)}})^{-1}]^{-1}\,,
\end{equation}
and in case of the relativistic protons we use
\begin{equation}
    \tau_{\mathsf{cool}}^{\rm (p)}=[(\tau_{\mathsf{syn}}^{\mathsf{(p)}})^{-1}+(\tau_{\mathsf{C}}^{\mathsf{(p)}})^{-1}+(\tau_{\mathsf{p\gamma}}^\pi)^{-1}+(\tau_{\mathsf{BH}})^{-1}+\tau_{\mathsf{pp}}^{-1}]^{-1}\,,
\end{equation}
including the photopion ($\pi$), Bethe-Heitler pairs (BH), and hadronic pion (pp) production loss rates. Proton synchrotron losses---as well as the associated radiation---are negligible for the considered environments. Note that these processes require additional information on the associated interaction medium, which is one of the following targets:
\begin{enumerate}
    \item[(i)] A magnetic field, which is assumed to be uniform on small scales (with respect to the particles' gyro radius) and randomly orientated on significantly larger scales (due to isotropic Alfv\'enic turbulence).
    \item[(ii)] A photon target, which is in the case of the starburst zone dominated by the thermal IR emission due to the re-scattered starlight by dust grains with a temperature $\theta_{\rm dust}$ and can be described by an isotropic, diluted modified blackbody radiation field 
    \be
    n_{\rm IR}(E) = \frac{C_{dil}}{ \pi^2\,(\hbar c)^3}\,\frac{E^2 }{\exp(E/(k_{\rm B}\,\theta_{\rm dust}))-1}\,\left( \frac{E}{E_0}  \right)\,,
    \label{eq:diff_IR-photon-dens}
    \ee
    where the dust clouds become optically thick above a critical energy $E_0=8.2\,\text{meV}$ \citep{YunCarilli2002}. The constant dilution factor $C_{dil}$ is determined from the observed IR luminosity $L_{\rm IR}$ according to the relation $L_{\rm IR}/(\pi R_{\rm str}^2c)=\int\text{d} E\,\,E\,n_{\rm IR}(E)$.\footnote{A more accurate approach of the IR photon spectrum has been proposed by \cite{Casey2012}, where a coupled modified greybody plus a mid-infrared power law has been used, but these modifications have no impact on our results.}
    In case of the coronal region we used a parametrized model \citep{Ho2008} above $1\,\text{eV}$ that accounts for the optical and UV emission by the disk as well as the Comptonized X-ray emission by hot thermal electrons in the corona. Hereby, the parametrization depends on the Eddington ratio ($L_{\rm bol}/L_{\rm Edd}$), i.e.\ the ratio of the bolometric over the Eddington luminosity, and we adopt the relation of \cite{Hopkins+2007} to determine $L_{\rm bol}$ based on the intrinsic X-ray luminosity $L_{\rm X}$ between $(2-10)\,\text{keV}$.
    \item[(iii)] A thermal gas target with a given temperature $\theta$, which is due to mathematical convenience assumed to be homogeneously distributed in both regions. For the starburst ring \cite{Spinoglio+2012} determine $\theta=127\,\text{K}$, a gas density of $n(H_2) = 10^{2.9}\,\text{cm}^{-3}$, and a molecular hydrogen mass of $M(H_2) \sim 3.5 \times 10^8\,M_\odot$
\end{enumerate}
More details on the individual energy-loss timescales can be found in the Appendix \ref{sec:energy_losses}. 

In addition, we have to account for catastrophic particle losses according to the escape of particles from the considered zones in a total time $\tau_{\mathsf{esc}}$. Here, the total escape rate via gyro-resonant scattering through turbulence with a power spectrum $\propto k^{-\varkappa}$ and a turbulence strength $\eta^{-1}$ as well as a bulk stream flow can be approximated by \citep{a5:Murase+2020}
\begin{equation}
    \tau_{\mathsf{esc}}\simeq\begin{cases}
    \left[ \frac{\eta}{9}\,\frac{c}{R}\left( \frac{e\,B\,R}{T} \right)^{\varkappa-2} + \frac{v_{\rm w}}{R} \right]^{-1}\,,\quad&\text{for the starburst,}\\
    \left[ \frac{\eta}{9}\,\frac{c}{R}\left( \frac{e\,B\,R}{T} \right)^{\varkappa-2} + \frac{a_{\rm vis} v_{\rm K}}{R} \right]^{-1}\,,\quad&\text{for the AGN corona}\,,
    \end{cases}
\end{equation}
with respect to the characteristic size $R$ and the magnetic field strength $B$ of the zones. Here, $e$ denotes the elementary charge and $c$ refers to the speed of light. In the case of the starburst zone, the bulk motion is given by a galactic wind with a velocity $v_{\rm w}$; in the case of the AGN corona, we account for the infall timescale, which is expected to be similar to the advection dominated accretion flow \citep{a5:Murase+2020}, with a viscosity parameter $a_{\rm vis}$ and the Keplerian velocity $v_{\rm K}=\sqrt{G\,M_{\rm BH}/R}$. For the spectral index of the turbulence spectrum it is common to assume a value of either $\varkappa=5/3$ referring to Kolmogorov turbulence, or $\varkappa=3/2$ (Kraichnan turbulence) which can be motivated from isotropic MHD turbulence in the magnetically dominated regime. In the following, we will adopt Kolmogorov turbulence for both regions unless stated otherwise. 
As discussed by several previous works, e.g.~\cite{Kheirandish+2021, Inoue+2019}, the stochastic diffuse acceleration (SDA) in the coronal region typically appears to be inefficient compared to the cooling rates. Therefore, it has been suggested that there needs to be some other acceleration mechanism such as magnetic reconnection at work. However, there is very little known about the actual acceleration efficiency of this process in the AGN corona as well as the resulting spectral shape of the CR energy distribution, so that we choose to stick to the SDA process at first. Hereby, we adopt the ansatz that the same scattering process that yields the diffusive escape is also responsible for the stochastic acceleration. In the case of the starburst region, the acceleration is expected to be introduced by a multitude of supernova remnants (SNRs) via diffusive shock acceleration (DSA). Here, the CRs are kept within the accelerating shock region by the Bohm diffusion---which gets introduced by the CR self-generated Bell instability \citep{Bell2004}---where the wave turbulence typically inherits the flat energy spectrum of the generating cosmic rays, i.e. $\varkappa=1$. Further, we assume that the average shock speed $v_{\rm sh}$ equals the wind speed $v_{\rm w}$.
Thus, we use the acceleration timescale \citep[e.g.][]{a5:Murase+2020, a5:Romero+2018}
\begin{equation}
    \tau_{\mathsf{acc}}\simeq\begin{cases}
    \frac{20}{3}\,\frac{T}{e\,B\,c}\,\left(\frac{c}{v_{\rm sh}}\right)^2,\quad&\text{for the starburst,}\\
    \eta\,\left(\frac{c}{v_{\rm A}}\right)^2\,\frac{R}{c}\left( \frac{e\,B\,R}{T} \right)^{\varkappa-2}\,,\quad&\text{for the corona}\,,
    \end{cases}
\end{equation}
and set the maximal CR energy $\hat{T}$ according to that value where the acceleration timescale exceeds the competing total loss timescale, i.e.~$\tau_{\mathsf{acc}}=[(\tau_{\mathsf{cool}})^{-1}+(\tau_{\mathsf{esc}})^{-1}]^{-1}$. 
%
\begin{figure}[htbp]
    \centering
    \includegraphics[width=0.54\linewidth]{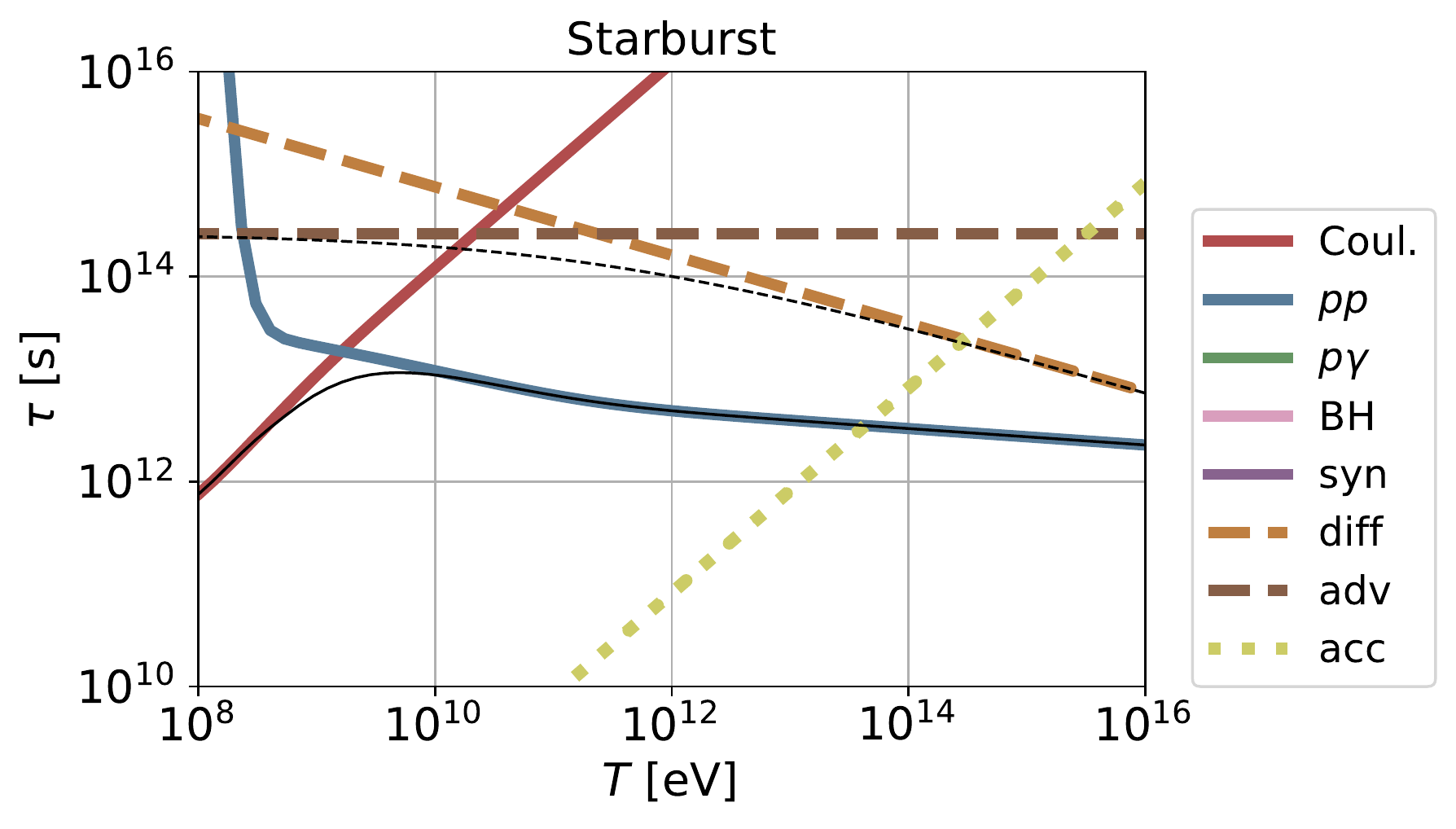}
    \includegraphics[width=0.445\linewidth]{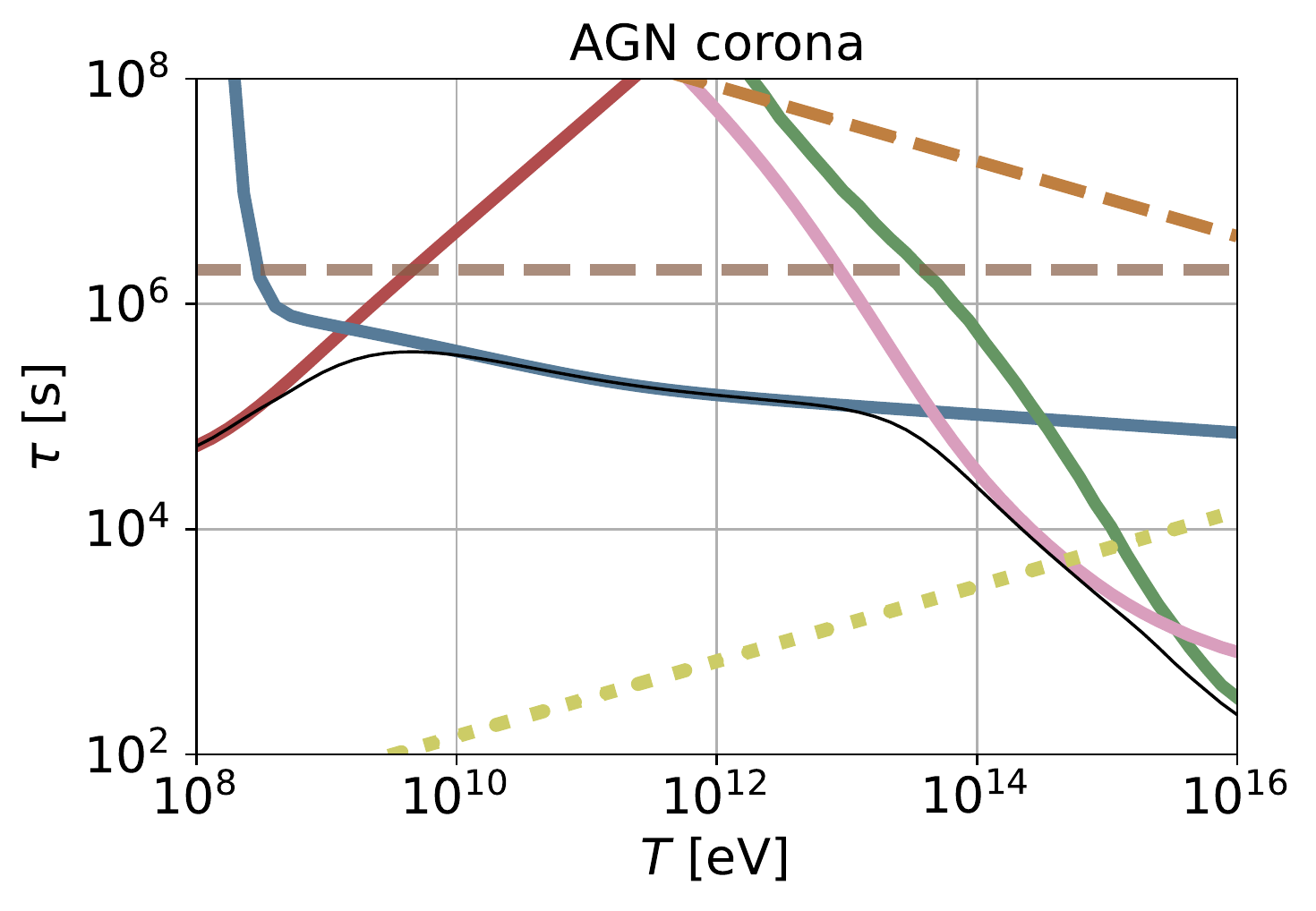}
    \includegraphics[width=0.54\linewidth]{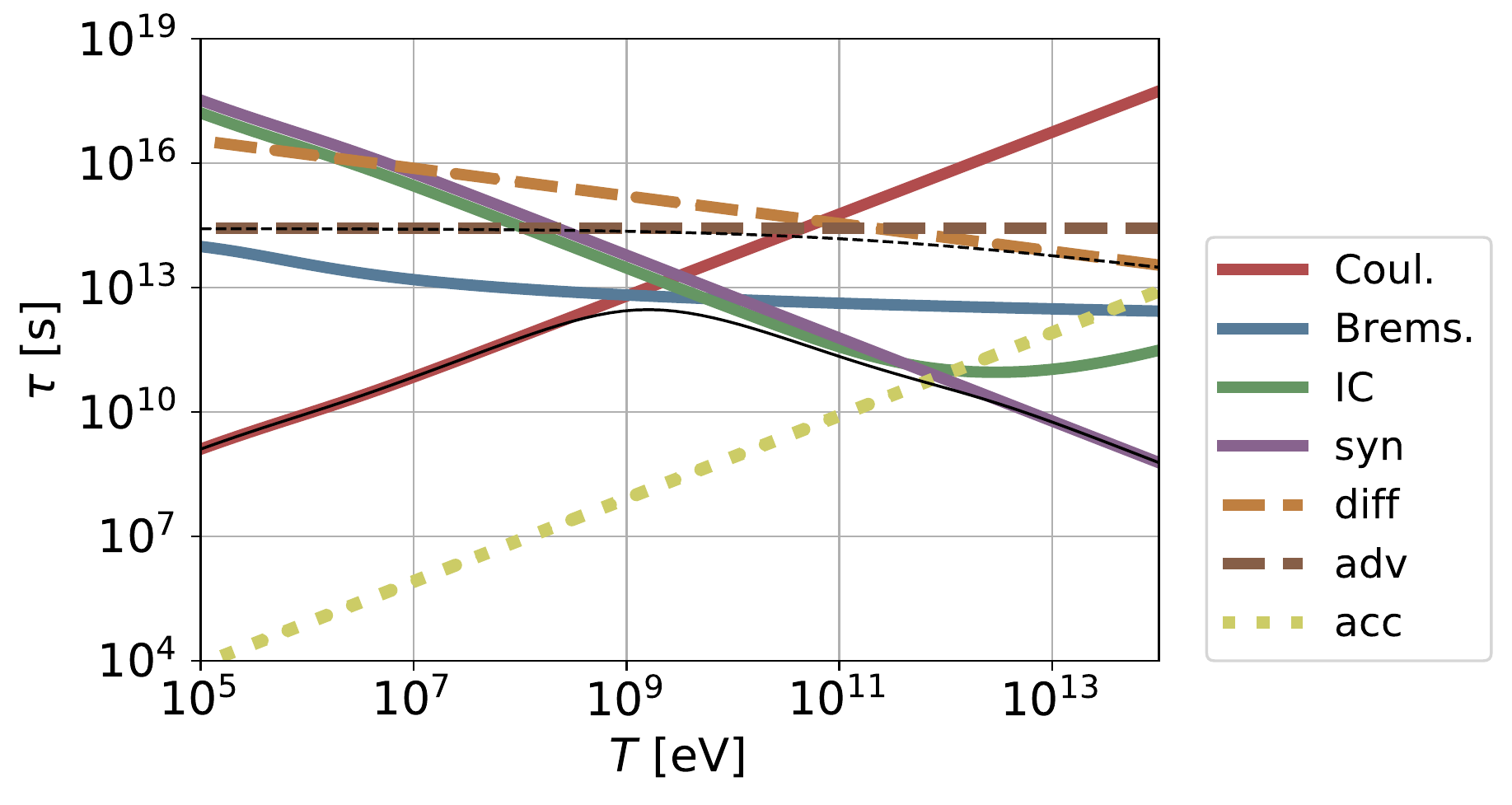}
    \includegraphics[width=0.44\linewidth]{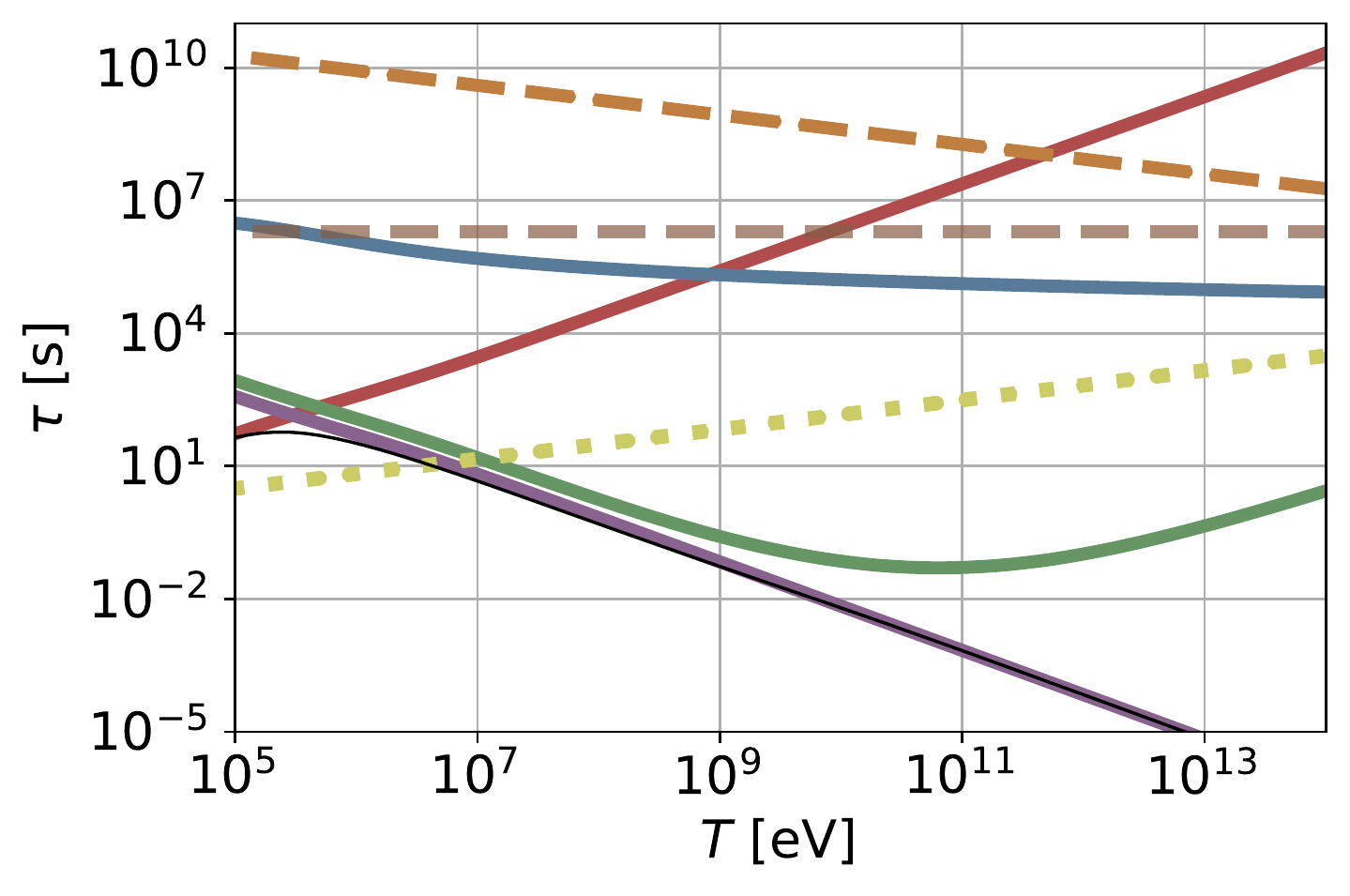}
    \caption{The different timescales of relativistic protons (\emph{upper panel}) and electrons (\emph{lower panel}) for the starburst zone (\emph{left}) and the AGN corona zone (\emph{right}). The assumed parameters are given in Table \ref{tab:Param}. The thin solid black line refers to the total energy loss timescale ($\tau_{\rm cool}$) and the thin dashed black line refers to the total escape timescale ($\tau_{\rm esc}$). The abbreviated individual process are as follows: $pp\hat{=}$hadronic pion production, $p\gamma\hat{=}$photopion production, BH$\hat{=}$Bethe Heitler pair production, syn$\hat{=}$synchrotron radiation, diff$\hat{=}$diffusion, adv$\hat{=}$advection, acc$\hat{=}$acceleration (DSA for the starburst and SDA for the corona), Coul.$\hat{=}$Coulomb losses, Brems.$\hat{=}$Bremsstrahlung, IC$\hat{=}$inverse Compton scattering.}
    \label{fig:timescales} 
\end{figure}
Note that due to the huge differences in the physical parameters---such as the gas density or the magnetic field strength---the different timescales differ significantly between the starburst and the corona zone, as shown in Fig.~\ref{fig:timescales}. 
With respect to the maximal CR energy $\hat{T}$ it is shown that in case of the starburst DSA can typically provide a maximal primary proton (electron) energy of several tens of TeV (hundreds of GeV). In the AGN corona a high turbulence strength ($\eta\sim 1$) or a rather flat turbulence spectrum ($\varkappa\lesssim 3/2$) is needed to obtain CR proton energies of about $100\,\text{TeV}$ or more which would be necessary to stay within the limits of the IceCube observations at the indicated potential flux \citep{a5:IceCube2020}. Still, the coronal CR electrons typically suffer from significant synchrotron/ IC losses at energies as low as a few tens of MeV. In addition, these primary electrons also have to overcome the Coulomb losses at low energies, and hence, need to be injected into the SDA process at about few keV.\footnote{Note that we do not account for the possible steepening of the turbulent power spectrum at energies below the thermal proton energy, which would lengthen the acceleration time considerably.} 
\subsection{Spectral energy distribution of CR electrons and protons} \label{sec:CR-SED}
\begin{figure}[htbp]
    \centering
    \includegraphics[width=0.49\linewidth]{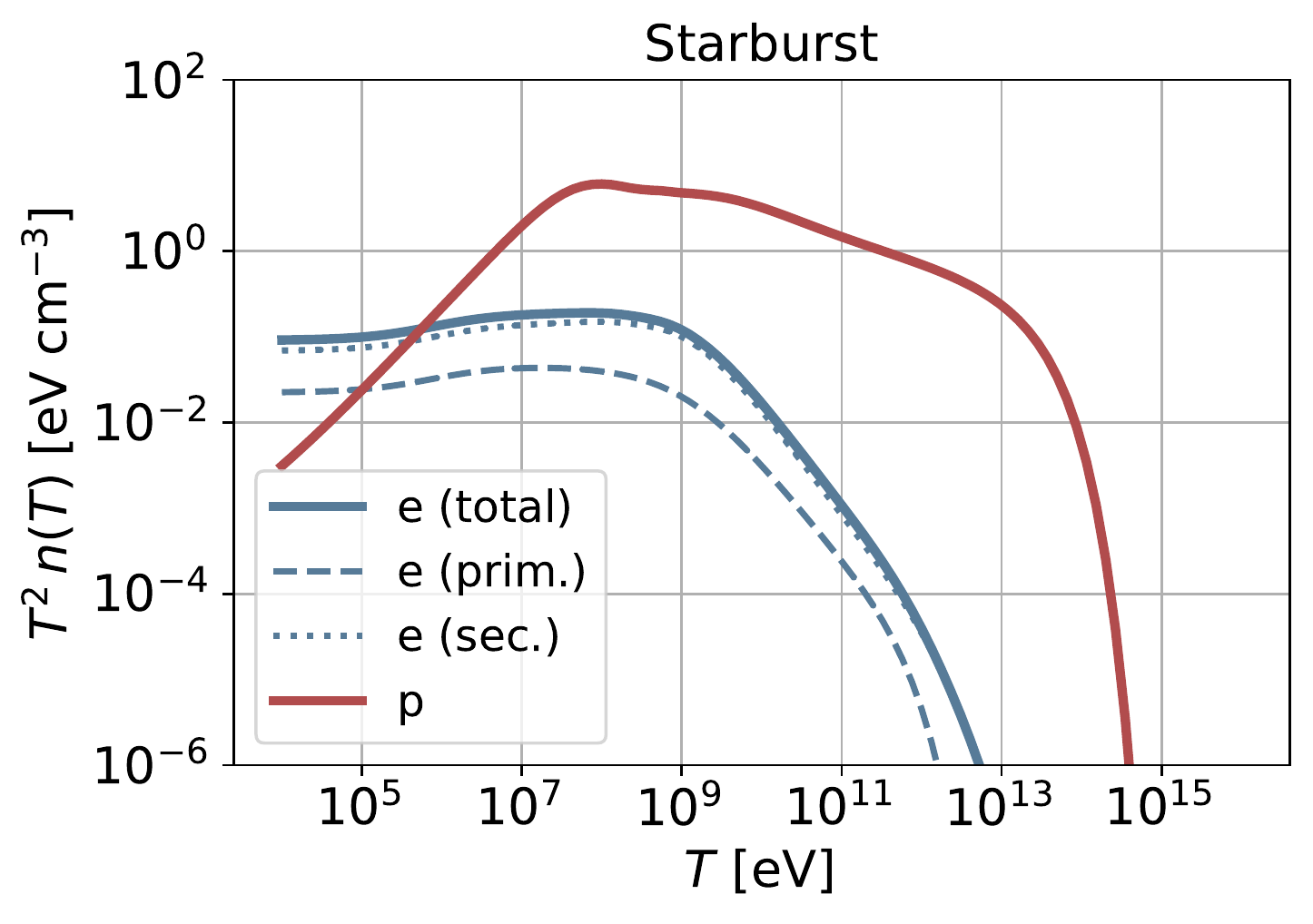}
    \includegraphics[width=0.49\linewidth]{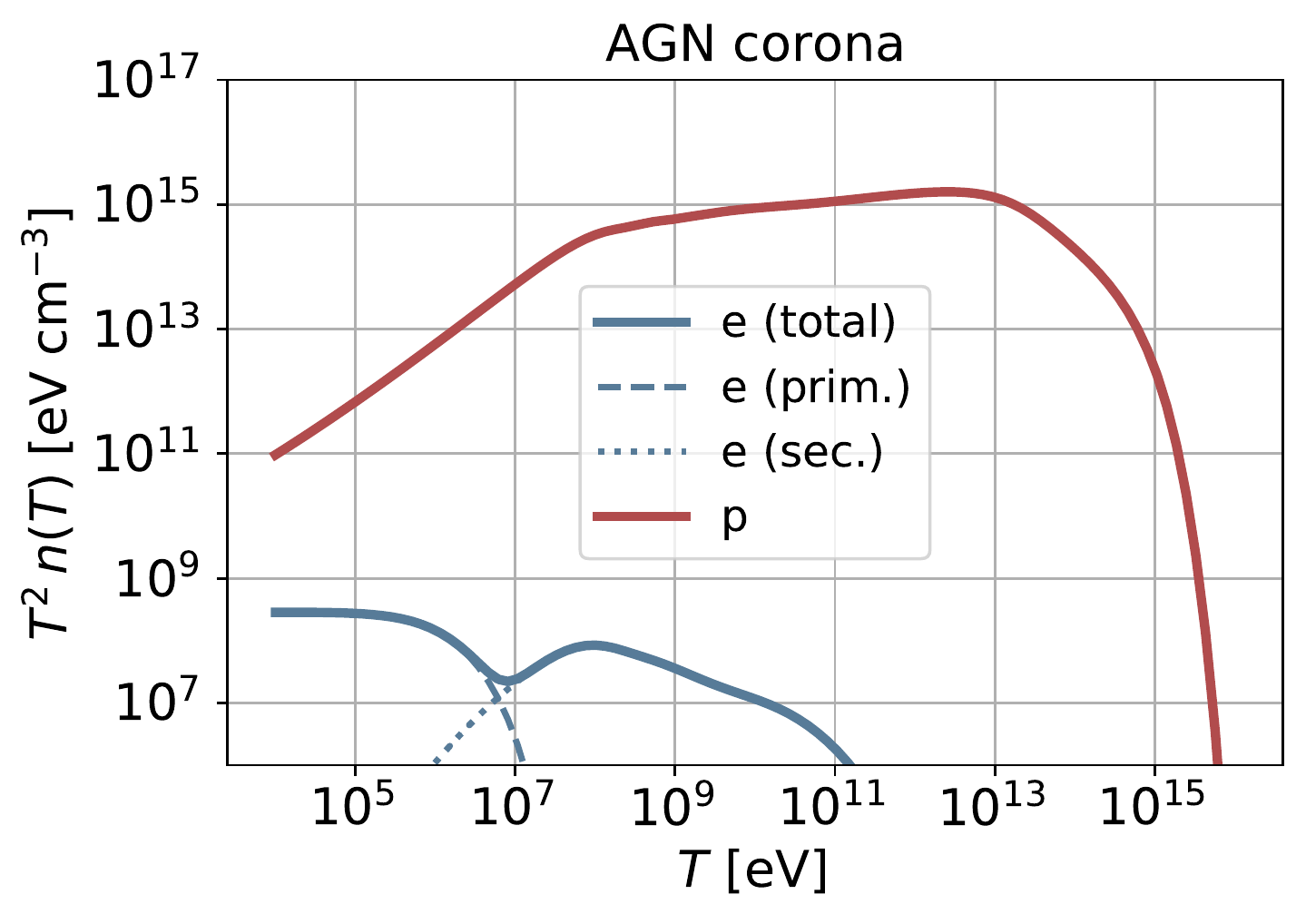}
    \caption{The resulting CR spectra of the starburst zone (\emph{left}) and the AGN corona zone (\emph{right}) using the best fit parameters as introduced in Sect.~\ref{sec:results}---hereby $s=2.2$ for the starburst and $s=1.7$ for the corona.}
    \label{fig:CRspectra} 
\end{figure}

Solving the transport equation (\ref{eq:teq0}) by fundamental methods (see  e.g.~\citealp{a5:eichmann2016}) provides the differential non-thermal electron and proton density $n(T)$ in the starburst as well as the coronal zone, as shown in Fig.~\ref{fig:CRspectra}. 
Despite the assumed quasi-neutrality of the primary source rates, the resulting fraction of CR protons is significantly higher than the one of the CR electrons, especially in case of the coronal region. This is due to significantly smaller energy loss time scales of the relativistic electrons according to synchrotron and IC losses (see Fig.~\ref{fig:timescales} ). Both regions are perfect calorimeters (except for CR protons of the starburst region with energies above about $1\,\text{PeV}$), so that the energy distributions steepen at certain characteristic energies either due to one of the energy loss processes or due to the exponential cut-off introduced by the acceleration. In Fig.~\ref{fig:CRspectra} we consider a parameter scenario that results from a fit to the data (as introduced in Sect.~\ref{sec:results}) and it can be seen that, for this particular scenario, the secondary electrons have a major contribution to the total CR electron spectrum. The individual contributions to its source rate $q^{\rm 2nd}_{\rm e}$ will be introduced in the following.

\subsection{Non-thermal emission} \label{sec:non-thermal_Emission}
Using the spectral energy distribution of CR protons and electrons we determine the non-thermal emission of those particles from radio to gamma-ray energies. Hereby, both emission regions are considered to be spherically symmetric, with the inner coronal region constrained to a radius $R_{\rm cor}=r_{\rm cor}\,\mathcal{R}_{\rm s}$, where typically $r_{\rm cor}\in [1,100]$ and $\mathcal{R}_{\rm s}=2GM/c^2$ denotes the Schwarzschild radius, and the surrounding starburst ring extends between an inner radius $R_{\rm str}^{\rm in}\sim 1\,\text{kpc}$  \cite[][]{Rico-Villas+2021} and an outer radius $R_{\rm str}^{\rm out}\in [1.3,2]\,\text{kpc}$. 

In the following we include the non-thermal photon emission by synchrotron radiation, inverse Compton scattering, non-thermal bremsstrahlung from the CR electrons according to a spectral emissivity $\epsilon_{\rm syn}$, $\epsilon_{\rm ic}$, and $\epsilon_{\rm brems}$, respectively, as well as hadronic and photo pion production by the CR protons with a spectral emissivity $\epsilon_{\rm pp}$ and $\epsilon_{\rm p \gamma}$, respectively. These hadronic processes also introduce high-energy neutrinos as well as secondary electrons and positrons with a source rate $q^{\rm 2nd}_{\rm e} = \epsilon_{\rm pp}^{\rm (e)}+\epsilon_{\rm p \gamma}^{\rm (e)} + \epsilon_{\rm BH} + \epsilon_{\rm \gamma\gamma}$, where the last two terms introduce electrons/positrons by $\gamma\gamma$ and Bethe-Heitler pair production. 

In addition to these non-thermal emission processes, we also account for the free-free emission ($\epsilon_{\rm ff}$) by the considered thermal target gas. Note that we use the spatially averaged thermal gas distribution in both zones. The actual gas distribution might differ. This would change the corresponding free-free emission. Further there are hints \citep{Inoue+2020} of a colder, less dense gas distribution in the extended coronal region which might contribute at radio/IR frequencies. But since there is no need for such an additional component in our model, we neglect the thermal emission by the extended coronal region to limit the number of free parameters.
The total spectral emission rate of the considered messenger particle $m=(\gamma,\nu)$ is given by
\be
\epsilon_m(E_m) = 
\begin{cases}
\epsilon_{\rm syn}(E_\gamma)+\epsilon_{\rm ic}(E_\gamma)+\epsilon_{\rm brems}(E_\gamma)+\epsilon_{\rm pp}^{\rm (\gamma)}(E_\gamma)+\epsilon_{\rm p \gamma}^{\rm (\gamma)}(E_\gamma)+\epsilon_{\rm ff}(E_\gamma)\,,\quad&\text{for } m=\gamma\\
\epsilon_{\rm pp}^{\rm (\nu)}(E_\nu)+\epsilon_{\rm p \gamma}^{\rm (\nu)}(E_\nu)\,,\quad&\text{for } m=\nu
\end{cases}
\ee

Moreover, it is necessary---in particular for the coronal region---to account for different absorption processes such as synchrotron-self absorption ($\alpha_{\rm syn}$) and free-free absorption ($\alpha_{\rm ff}$) at radio/IR energies as well as $\gamma\gamma$ pair production ($\alpha_{\gamma\gamma}$) at gamma-ray energies, yielding a total absorption coefficient 
\be
\alpha_m(E_m) = 
\begin{cases}
\alpha_{\rm syn}(E_\gamma)+\alpha_{\rm ff}(E_\gamma)+\alpha_{\gamma\gamma}(E_\gamma)\,,\quad&\text{for } m=\gamma\\
0 \,,\quad&\text{for } m=\nu
\end{cases}
\ee
Note that we consider both source regions as optically thin for the neutrinos, which does not necessarily need to be the case for the dense coronal region at the highest energies.
More details on the individual emissivities as well as the absorption coefficients can be found in the Appendix \ref{sec:emissivities} and \ref{sec:absorption_coeff}. 

For a homogeneous source for which $\epsilon_m$ and $\alpha_m$ are constants, the spectral energy flux at a source distance $d$ is generally given by \citep[e.g.][]{Gould1979}
\begin{equation}
    F_m(E_m) = \frac{E_m}{4\pi\,d^2}\,\epsilon_m(E_m)\,V_{\rm eff}(E_m)
\end{equation}
with the effective emission volume
\begin{equation}
    V_{\rm eff}(E_m) = \int \diff^3 r\,\exp\left( - \alpha_m(E_m)\, |\vec{r}_{\rm s}-\vec{r}|\right)\,,
\end{equation}
where $\vec{r}_{\rm s}$ represents an arbitrary position on the surface of the emission volume. Due to the different geometry of the considered emission sites also the effective emission volume differs. Considering the directional symmetries of both systems the previous equation can be simplified to 
\begin{equation}
    V_{\rm eff}(E_m) = 
    \begin{cases}
    \int_{0}^{2\pi}\int_{0}^{2\pi}\int_0^{a_-}\diff r\,\diff u\,\diff \nu \,\,\, r \left( a_{+}+r\,\cos\nu \right)\,\exp\left[ - \alpha_m(E_m)\, G_{\rm str}(r,u,\nu)  \right]\,,\quad&\text{for the starburst,}\\
    \frac{\pi\,R_{\rm cor}^2}{\alpha_m(E_m)}\left[ 1 + \frac{(2\alpha_m(E_m)R_{\rm cor}+1)\exp(-2\alpha_m(E_m)R_{\rm cor})-1}{2[\alpha_m(E_m)R_{\rm cor}]^2}  \right]\,,\quad&\text{for the corona}\,.
    \end{cases} 
\end{equation}
with 
\be
G_{\rm str}(r,u,\nu) = \sqrt{ 2a_+^2(1-\sin u) + 2a_+r\cos\nu(1-\sin u) + r^2(1+\sin^2\nu)+a_-^2-2a_-r\sin\nu}\,,
\ee
and $a_\pm=(R_{\rm str}^{\rm out}\pm R_{\rm str}^{\rm in})/2$. 
Note that for optically thin emission, i.e.\ in the case that $\alpha_m$ is significantly smaller than the characteristic length scales of the system, we obtain
\be
V_{\rm eff}(E_m) = 
    \begin{cases}
    2\,\pi^2\, a_-^2\,a_+\,,\quad&\text{for the starburst,}\\
    4\,\pi\,R_{\rm cor}^3/3\,,\quad&\text{for the corona}\,,
    \end{cases} 
\ee
as expected.

\section{Explaining the Multi-Messenger Data} \label{sec:results}
\subsection{Data and Fit Procedure}
In the following we use the previously introduced model to explain the multi-messenger data of NGC 1068. Hereby, we quantify the goodness of the fit by the chi-squared value $\chi^2=\sum_j (P_j-O_j)^2/\sigma^2(O_j)$, where $P_j$ ($O_j$) denotes the model prediction (observation) of the flux of photons and neutrinos at different energies. As the model prediction depends in general on a large parameter set (see Table~\ref{tab:Param}), we use the
differential evolution algorithm\footnote{\url{https://docs.scipy.org/doc/scipy/reference/generated/scipy.optimize.differential_evolution.html}} to find a global minimum within a dedicated subset of this parameter set. Hereby, we keep fixed those parameters that have either a minor impact on the resulting flux prediction---such as the wind speed $v_w$ and the characteristic temperature $\theta_{\rm dust}$ of the dust grains in the starburst region, the radiative efficiency of the disk, and the viscous parameter $\mu_{\rm vis}$ in the accretion flow (where we adopt the same value as suggested by \citealp{Schartmann+2010})---or are rather well defined from observations: 
we use the intrinsic $2-10\,\text{keV}$ X-ray luminosity $L_{\rm X}=10^{43}\,L_{43}\,\text{erg/s}$ of the coronal region as derived from two different analyses, as this luminosity carries large uncertainties due to the high column density of the source. Hence, a detailed modelling of the impact of the torus is necessary to obtain the actual fraction of X-rays that is scattered into the line of sight. In the energy range between 2 and 10\,keV NuSTAR and XMM-Newton monitoring campaigns \citep{Marinucci+2016} yield an intrinsic luminosity of $L_{43}=7^{+7}_{-4}$, whereas a different analysis by \cite{Ricci+2017} that combines 70-month averaged Swift/BAT data with different measurements in this soft X-ray band obtains $L_{43}=0.9$.

In general, it is expected \citep[e.g.][]{Mayers+2018_arxiv} that a higher X-ray luminosity corresponds to higher black hole mass $M_{\rm BH}$ than a low X-ray luminosity. As shown by different observations \citep[see][and references therein]{GRAVITY2020} the expected range of the black hole mass yields $M_{\rm BH}\in[0.8,\,1.7]\times 10^7\,M_\odot$ and since a higher X-ray luminosity generally indicates a higher value of $M_{\rm BH}$ \citep[e.g.][]{Mayers+2018_arxiv}, we use $M_{\rm BH}(L_{43}=0.9)=0.8\times 10^7\,M_\odot$ and $M_{\rm BH}(L_{43}=7)=1.7\times 10^7\,M_\odot$, respectively, yielding an Eddington luminosity of $L_{\rm Edd}(L_{43}=0.9)=1.0\times 10^{45}\,\text{erg s}^{-1}$ and $L_{\rm Edd}(L_{43}=7)=2.1\times 10^{45}\,\text{erg s}^{-1}$, respectively. And based on the relation by \cite{Hopkins+2007} we obtain a bolometric luminosity of $L_{\rm bol}(L_{43}=0.9)=0.2\times 10^{45}\,\text{erg s}^{-1}$ and $L_{\rm bol}(L_{43}=7)=2.9\times 10^{45}\,\text{erg s}^{-1}$, which is nicely within the range that has been found by others \citep[see][and references therein]{GRAVITY2020}. 
Hence, this also has a direct consequence on the adopted black hole properties, so that the Schwarzschild radius $\mathcal{R}_{\rm s}$ as well as the mass accretion rate $\dot{M}$ decrease for a decreasing $L_{\rm X}$ value.
The surrounding starburst region is dominant in the IR, due to scattering off of dust grains with a characteristic temperature $\theta_{\rm dust}$. 
In the following we adopt $\theta_{\rm dust}=127\,\text{K}$ as observed by \cite{Spinoglio+2012}, although also higher temperatures of up to 250 K have been observed recently \citep{Rico-Villas+2021}. But as previously mentioned the impact of $\theta_{\rm dust}$ on our results is negligible. Further, we use the observed bolometric mid-infrared luminosity $L_{\rm IR}$ of the outer starburst ring of about $30^{\prime\prime}$ in diameter, which is found to account for almost half its total mid-infrared luminosity \citep{Bock+2000}.

But also for the other, non-fixed parameters we account for physical constraints based on observations or numerical simulations. For example the optical depth of the coronal region is typically $\omega_{\rm T}\sim 0.1-1$ \citep{Ricci+2018,MerloniFabian2001} constraining the gas density according to $n_{\rm gas}=\omega_{\rm T}/(\sigma_{\rm T}R)$ if the coronal plasma is not dominated by electron-positron pairs. And based on the virial gas temperature $\theta_{\rm gas}=GM_{\rm BH}m_{\rm p}/(3R_{\rm cor}k_{\rm B} )= m_pc^2/(6r_{\rm cor}k_{\rm B} )$ of the protons we are able to draw constraints on the coronal magnetic field strength according to $B=\sqrt{8\pi n_{\rm gas} k_{\rm B} \theta_{\rm gas}/\beta}$, where a low plasma beta ($\beta\sim0.1-3$) is expected from numerical MHD simulations \citep{Jiang+2019,Jiang+2014,MillerStone2000}. All of the subsequently used parameter values and the constraints, respectively, are summarized in Table~\ref{tab:Param}. 
\begin{table}[ht]
\begin{center}
\caption{Summary of fixed and free parameters that describe the radiation zones of NGC 1068 in our model. The top lines give the parameters that describe the starburst region, namely the spectral index of the electron and proton distributions, $s$, the fraction of supernova energy that gets converted to CRs, $f_{\rm SN}$, the gas density, $n_{\rm gas}$, the magnetic field strength, $B$, the radius of the starburst ring, $R_{\rm str}$, the turbulence strength parameter, $\eta$, the spectral index of the turbulence spectrum, $\varkappa$, the minimal kinetic energy of the CR particles, $T_{\rm inj}$, the average wind speed, $v_{\rm w}$, and the infrared luminosity, $L_{\rm IR}$. The bottom two lines give the parameters that describe the AGN environment which include some of the starburst related parameters and in addition, the fraction of X-ray emissivity that goes into relativistic protons, $f_{\rm inj}$, the radius of the corona, $R_{\rm cor}$, the viscous parameter, $\mu_{\rm vis}$, and the intrinsic coronal X-ray luminosity, $L_{\rm X}$. For the free parameters, the scanning range is reported in the table.}
{ \footnotesize
\begin{tabular}{ccccccccccc}
    \multicolumn{11}{c}{Starburst}\\
    \cline{1-11}
$s$ & $f_{\rm SN}$ & $n_{\rm gas}$ [\text{cm}$^{-3}$] & $B$ [$\mu$G] & $\eta$ & $R_{\rm str}^{\rm out}$ [kpc] & $R_{\rm str}^{\rm in}$ [kpc] & $\varkappa$ & $T_{\rm inj}$ [MeV] & $v_{\rm w}$ [km/s] & $L_{\rm IR}$ $[10^{11}L_\odot]$ \\ 
\cline{1-11}
$1.5-2.5$ & $0.01-0.25$ & $10-1000$ & $10-500$ & $1-100$ & $1.3-1.8$ & $0.7$ & $5/3$ & $0.01$ & $100$ & 1.5\\
\hline
\end{tabular}
\vspace{0.5cm}\\
\begin{tabular}{cccccccccc}
    \multicolumn{10}{c}{AGN corona}\\
    \cline{1-10}
    $s$ & $f_{\rm inj}$ & $n_{\rm gas}$ [$10^9\text{cm}^{-3}$] & $B$ [kG] & $\eta$ & $R_{\rm cor}$ [$\mathcal{R}_{\rm s}$] & $\varkappa$ & $T_{\rm inj}$ [MeV] & $\mu_{\rm vis}$ & $L_{\rm X}$ $[10^{43}\text{erg/s}]$ \\ 
\cline{1-10}
$1.5-2.5$ & $0.00001-0.1$ & $0.05-50$ & $0.001-10$ & $1-100$ & $5-500$ & $5/3$ & $0.01$ & $0.1$ & $(0.9,\,7)$\\
\hline
\label{tab:Param}
\end{tabular}}
\end{center}
\end{table}
The observational data that were used in the fitting procedure are as follows:
\begin{enumerate}
    \item[(i)] The radio data \emph{of the starburst region} have first been observed by \cite{WilsonUlvestad1982}, who could, however, only disentangle the large-scale emission at 20\,cm. Therefore, \cite{Wynn-Williams+1985} re-observed the large scale emission of that source with VLA at 2, 6 and 20\,cm using $6^{\prime\prime}{.}5$ FWHM beams. Based on the given range of the flux density of $(4-8)\,\text{mJy per beam}$ at 2\,cm as well as the spectral behavior, we obtain an integral flux of $0.10\pm 0.03$, $0.31\pm 0.10$ and $0.81\pm 0.27$ Jy at 2, 6 and 20\,cm, respectively. Here, we consider the spatial region at a distance between $10^{\prime\prime}$ and $25^{\prime\prime}$ with respect to the nucleus due to the observed plateau in the flux density. Using the latest distance measurements---so that $1^{\prime\prime}\sim 70\,\text{pc}$---this large scale emission arises from a size between about 0.7 and 1.8\,kpc. The resulting flux at 20\,cm is only smaller by a few percent than what has been found earlier by \cite{WilsonUlvestad1982}, although these authors considered distances up to $2^{\prime}$. Further, both works conclude that the origin of this emission is almost certainly synchrotron emission. We are not aware of any additional recent data of this spatial region in the radio band, which could be used to constrain our starburst model at low energies. 
    \item[(ii)] The radio and IR data \emph{of the coronal region}, require a high spatial resolution of the observational instrument to exclude any additional contribution, e.g.\ by the torus which emits predominantly in the IR. Two components have been identified within the central parsecs of that source. One of those is the compact $(0.5-1.4)\,\text{pc}$ sized core of the AGN which we will associate with the coronal region in the following. Still these spatial scales are at least three order of magnitude larger than the inner corona, that we are modeling here. Thus, there is the chance that the so-called \emph{extended corona region} at distance of about mpc to pc from the black hole, provides an additional contribution. Previous works \citep{Roy+1998,Gallimore+2004,Inoue+2020} have shown that free-free emission by a gas with an electron temperature of a few $\times 10^6 \,\text{K}$ of this extended corona is actually able to explain the VLBA radio observation. Despite the recent objections (which we will discuss in more detail in Sect.~\ref{sec:conclusions}) by \cite{BaskinLaor2021}, we adopt this simple approach using a gas with an (electron/proton) density of $2.5\times 10^5\,\text{cm}^{-3}$ and an electron temperature of $10^6\,\text{K}$ to explain the resolution matched flux densities \citep{Gallimore+2004} of $<0.7$, $5.9\pm0.5$ and $5.4\pm0.5$ mJy at $1.4$, $5$ and $8.4\,\text{GHz}$, respectively (see Fig.~\ref{fig:SED_details}). Note, that the assumed gas becomes optically thick at about $5\,\text{GHz}$ which introduces a slight tension with the observed flux at that frequency, but yields a flux at $1.4\,\text{GHz}$ that is in agreement with the upper limit. 
    To explain the steep flux increase in the IR, however, an additional contribution is needed which we suggest to be given by the inner corona. At these frequencies the ALMA observatory has determined a flux density of $6.6 \pm 0.3$ and $13.8 \pm 1.0 \,\text{mJy}$ at 256 and 694\,GHz, respectively, using a beam sizes of 20 and 60 mas, respectively \citep{Impellizzeri+2019, ALMA2016}. In addition, we also include the recent results of the data analysis of the continuum fluxes at 224, 345, and 356 GHz with a beam sizes of 30 mas \citep{Inoue+2020}. The different resolutions of these observations introduce a mismatch in particular if an additional, strong flux contribution by the torus emerges at about a few tens of pc. Hence, the flux prediction at 224 GHz is expected to become smaller then the one from \cite{Impellizzeri+2019} at 256\,GHz, if the mismatched coverage would be taken into account properly. So, we account heuristically for this effect by introducing an additional, lower uncertainty $\Delta \check{F}_{\rm add}=(1-20\,\text{mas}/\delta_{\rm bs})\,F_{\rm obs}$ of the observed flux $F_{\rm obs}$ dependent on the beam size $\delta_{\rm bs}$.\footnote{Note that this additional uncertainty reduces the resulting $\chi^2$-value of the fit, but hardly affects the resulting best-fit parameters.}
    \item[(iii)] The gamma-ray data are taken up to $100\,\text{GeV}$ from the fourth Fermi-LAT catalog of gamma-ray sources \citep{4FGLcat2022}, and at higher energies we include the upper limits from the MAGIC telescope \citep{MAGIC2019}. At these frequencies one cannot resolve individual spatial regions and the data need to be explained by the total flux of both regions.
    \item[(iv)] The high energy neutrino flux that corresponds to the $2.9\,\sigma$ excess observed by the IceCube Neutrino Observatory is taken from Figure 7 of \cite{a5:IceCube2020}. In terms of the chi-squared calculation we only account for the most well constrained flux value at about $1\,\text{TeV}$ as well as flux at $28\,\text{TeV}$ to account for the steep spectral behavior.
\end{enumerate}
Note that we do not account for the flux attenuation from the coronal region by the torus, which is mostly relevant for $\lesssim 1\,\text{MeV}$. Here detailed modelling by \cite{Ricci+2017} of the torus absorption\footnote{Including the combined effect of photoelectric absorption and Compton scattering by neutral material as well as the absorption by ionized gas using the ZXIPCF model \citep{Reeves+2008}.} with respect to the broadband X-ray characteristics has shown that in the soft (hard) X-ray at $2-10\,\text{keV}$ ($14-195\,\text{keV}$) the coronal flux gets attenuated by a factor $0.018$ ($0.17$). 
Since the corona is a perfect CR calorimeter, as indicated by Fig.~\ref{fig:timescales}, there is no additional (hadronic or leptonic) emission by CR interactions in the torus region, so that we can completely neglect this region in the following. In addition to the inner corona, we also account for the free-free emission by the outer corona, that extends up to about $1\,\text{pc}$. As previously mentioned in (ii), its parameters are not changed by the fit algorithm, but fixed to explain the compact radio data.

\subsection{Fit Results}
Using the $(2\times6)$-dimensional parameter space of constrained fit parameters---as introduced by the first six columns in Table \ref{tab:Param}---we obtain a robust global chi-squared minimum of $\mathrm{min}(\chi^2)\simeq 8\,\,(10)$ for an intrinsic coronal X-ray luminosity of $L_X=7\,(0.9)\,\times 10^{43}\,\text{erg/s}$. Even though the resulting best-fit spectra are almost equal for both of those two cases---except for the resulting neutrino flux at $\lesssim 1\,\text{TeV}$---the resulting best-fit parameter space of the corona shows some differences: 
For $L_X=0.9\times 10^{43}\,\text{erg/s}$ the inner corona needs to extend about $(150-200)\,\mathcal{R}_{\rm s}$, which however, is about the same absolute size as for the case of a high X-ray luminosity. To further obtain a sufficient amount of CRs, a higher value of $f_{\rm inj}$ is needed due to the comparably small mass accretion rate. In addition the initial CR spectrum in the corona needs to be slightly softer ($s\sim 2$) to explain the IR and neutrino data, due to the smaller loss rate by IC scattering as well as photopion and Bethe-Heitler pair production.
But the rest of the resulting parameter space is similar to what is described in the following, in particular for the starburst ring. 

\begin{figure}[htbp]
    \centering
    \includegraphics[width=0.47\linewidth]{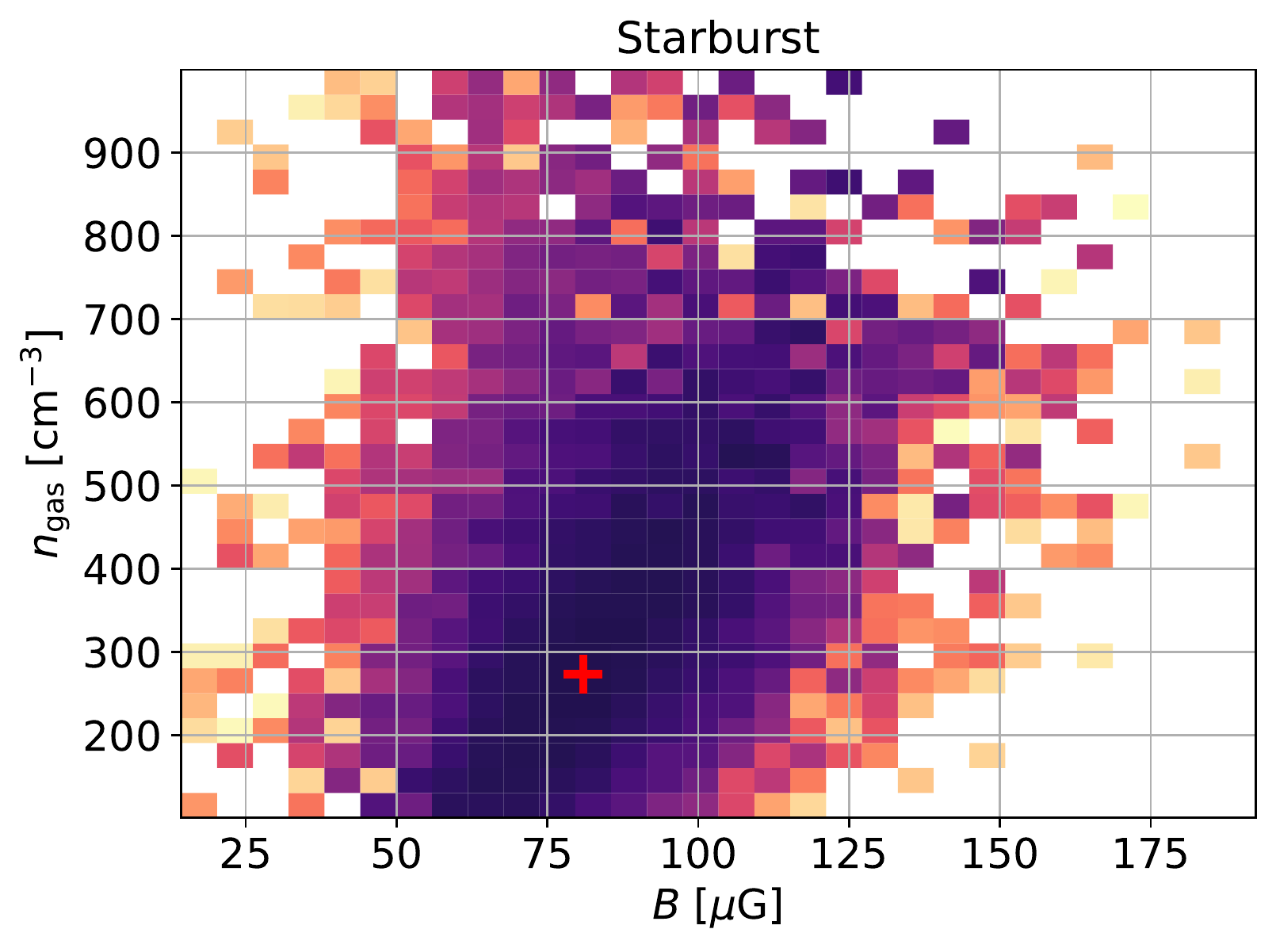}
    \includegraphics[width=0.515\linewidth]{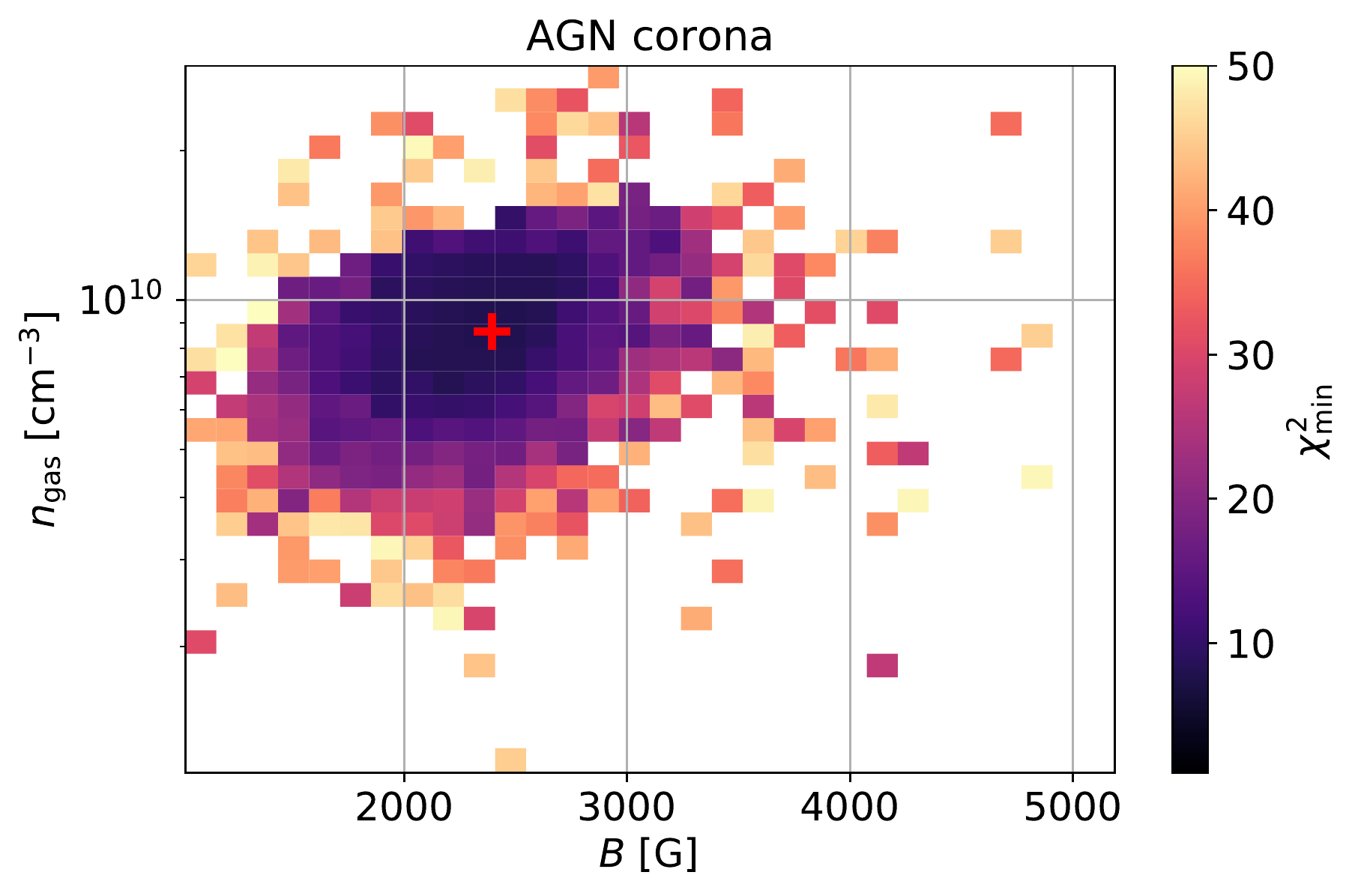}
    \includegraphics[width=0.47\linewidth]{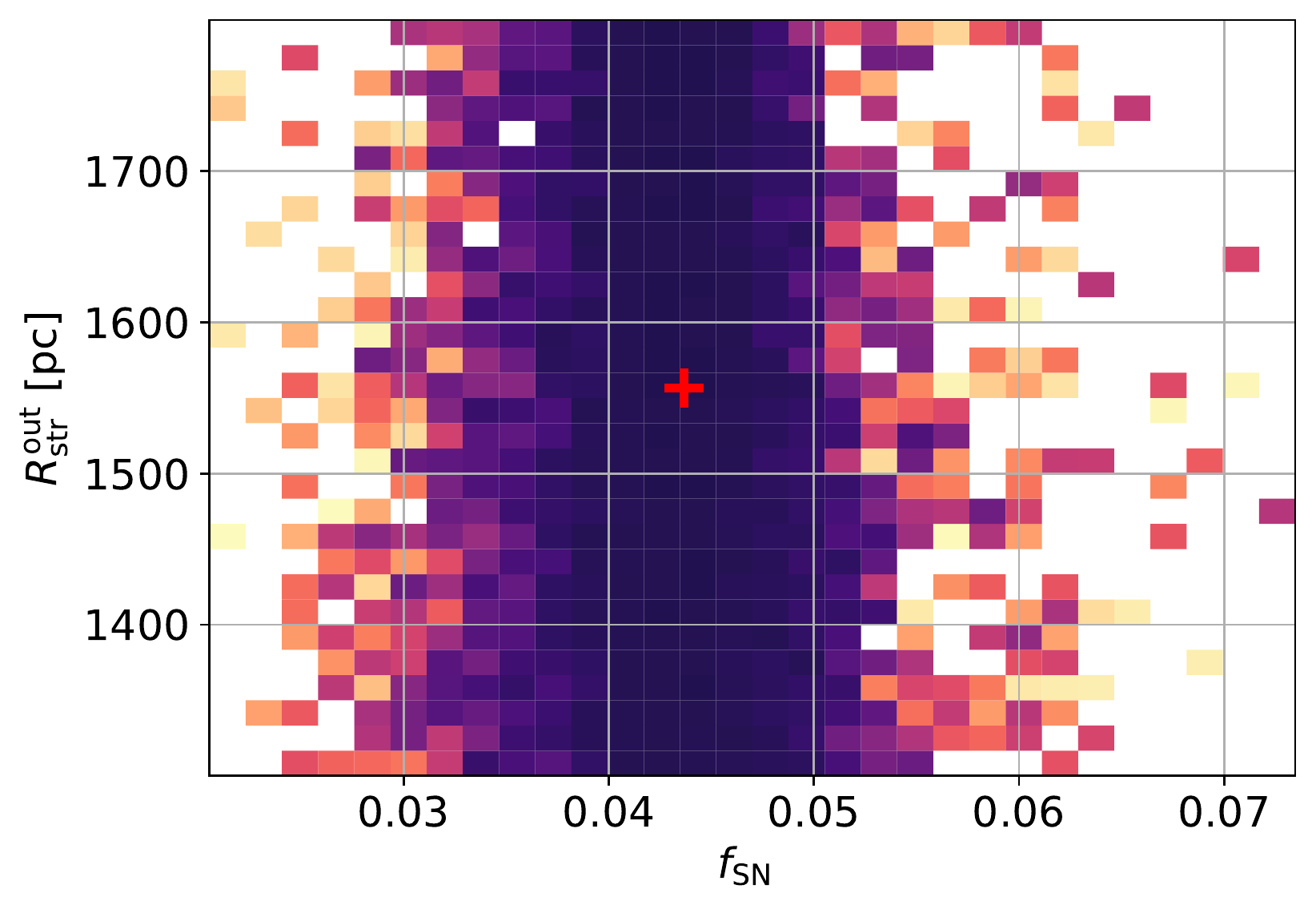}
    \includegraphics[width=0.515\linewidth]{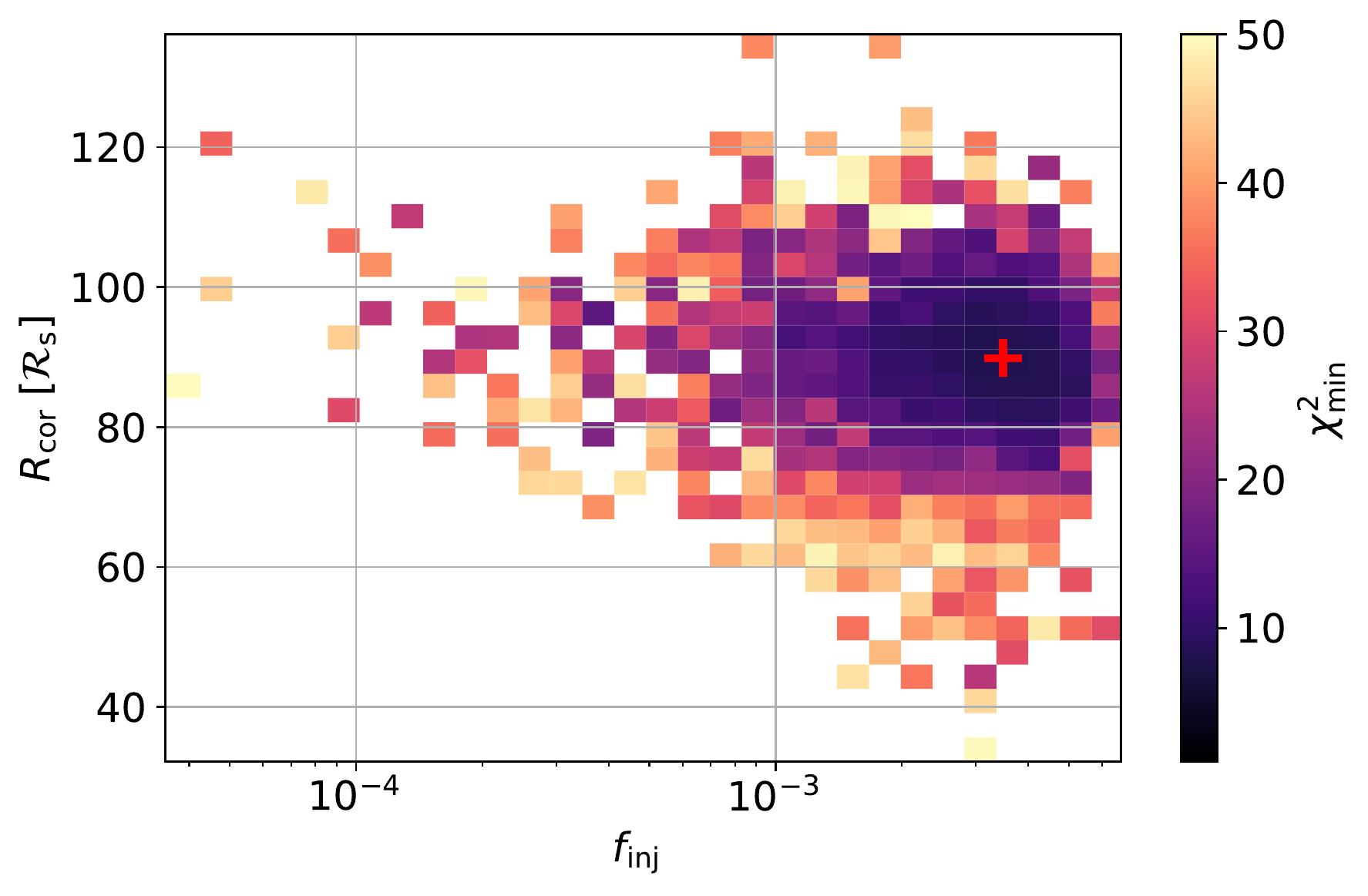}
    \includegraphics[width=0.47\linewidth]{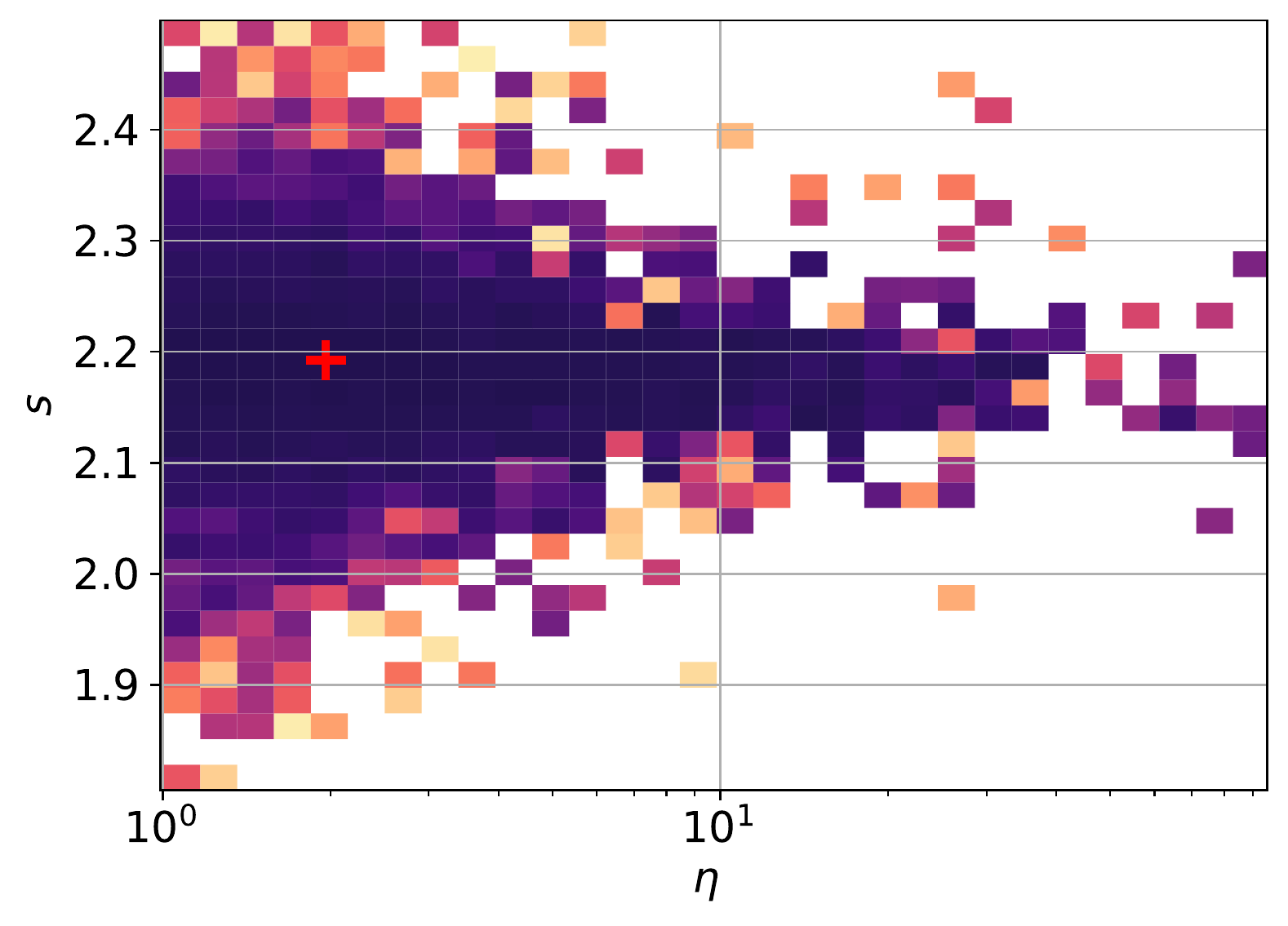}
    \includegraphics[width=0.515\linewidth]{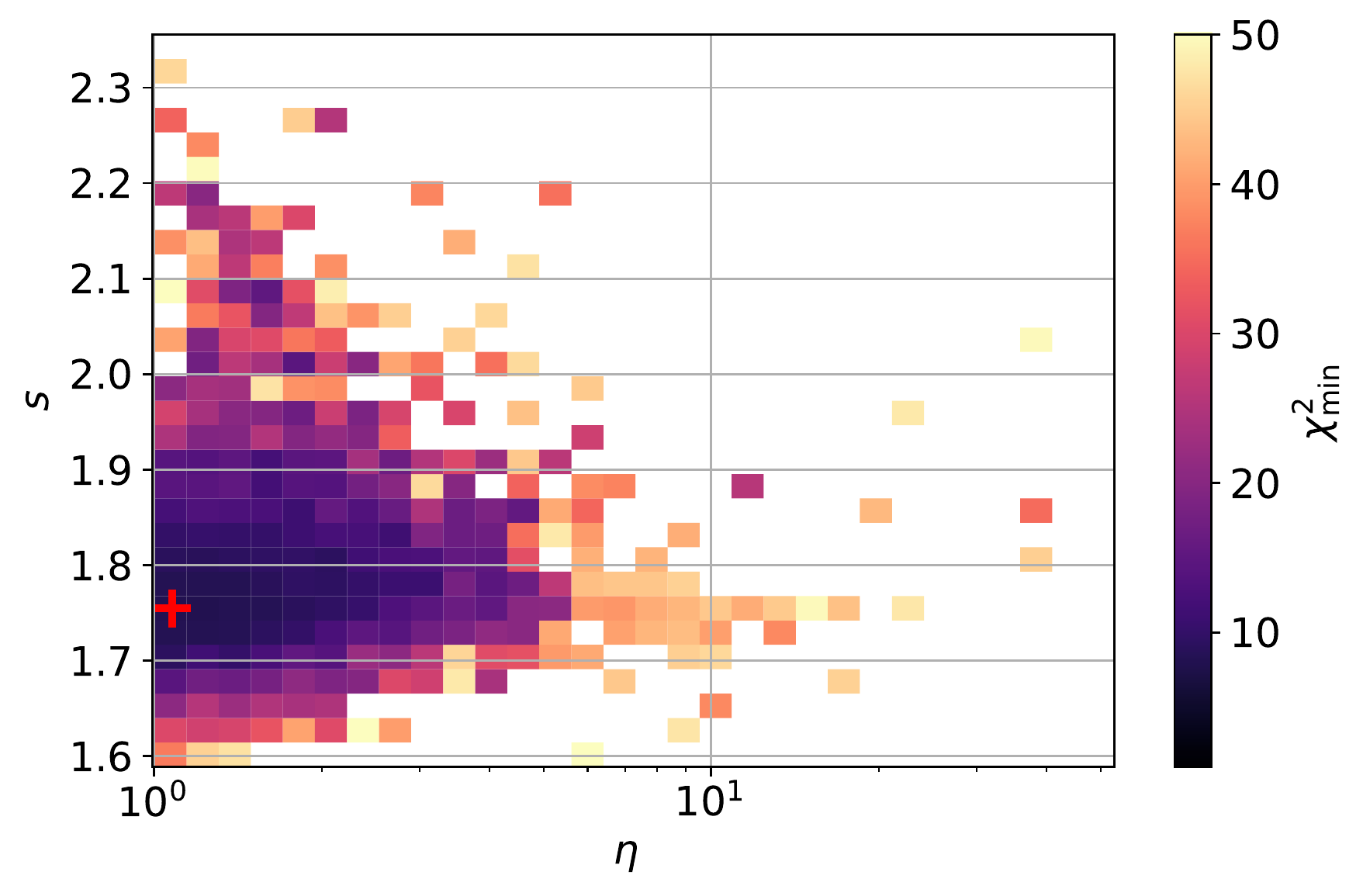}
    \caption{The chi-squared distribution of the starburst zone (\emph{left}) and the AGN corona zone (\emph{right}) dependent on the fit parameters, where we display for two different parameters and marginalise over the others. The red cross marks the best-fit parameter values that are also used in Fig.~\ref{fig:timescales}, \ref{fig:CRspectra} and \ref{fig:SED_details}.}
    \label{fig:chi-squared-results} 
\end{figure}

In the case of $L_X=7\times 10^{43}\,\text{erg/s}$, the Fig.~\ref{fig:chi-squared-results} shows the goodness of the fit for a certain range of the parameter space, where the chosen evolution strategy has converged. Here, the so-called \emph{'best1bin'} strategy is used, where two members of the population are randomly chosen and the difference is used to mutate the best member. Hence, the algorithm tends to increase the number of chi-squared function evaluation if the algorithm converges towards its minimum. An extended minimum in the parameter space is, in general, still favored with respect to a narrow one, so that it cannot be excluded, especially in such a multi-dimensional parameter space, that the resulting minimum is actually not a global but a local one. Since these are two-dimensional representations of the ($2\times 6$)-dimensional parameter space, we have to marginalise over the other dimensions, which is done by using its minimal chi-squared value $\chi^2_{\rm min}$.
A systematic scan of the whole parameter space would be needed to expose all of the details of the $\chi^2$-distribution. Still we can conclude that almost all of the 12 fit parameters have a significant impact on the goodness of the fit and the best-fit parameters (with $\chi^2\lesssim 10$) can be well constrained. However, this does not apply to the outer radius $R_{\rm str}^{\rm out}$ of the starburst ring, to which the fit results are not sensitive. Further, this best-fit parameter range does not represent any extreme scenarios, however, the inner corona needs a rather high gas density and magnetic field strength which extends up to about $(80-100)\, \mathcal{R}_{\rm s}$---note that the observed black hole mass of NGC~1068 is rather small compared to other AGN yielding a small Schwarzschild radius, so that $100\, \mathcal{R}_{\rm s}\simeq 1.5\times 10^{-4}\,\text{pc}$. Still, in combination with a high gas density of $\sim 10^{10}\,\text{cm}^{-3}$ this yields a rather high value of the optical Thomson depth of $\omega_{\rm T}\gtrsim 1$. In addition, these fit results suggest a strongly magnetized inner corona with a plasma beta $\beta\ll 1$. 
In case of the starburst region a high gas density of $n_{\rm gas}\sim 800\,\text{cm}^{-3}$, as suggested by \cite{Spinoglio+2012}, yields good agreement with the data, even though the best-fit result is obtained for a gas density that is about a factor of three smaller. 

\begin{figure}[htbp]
    \centering
    \includegraphics[width=0.99\linewidth]{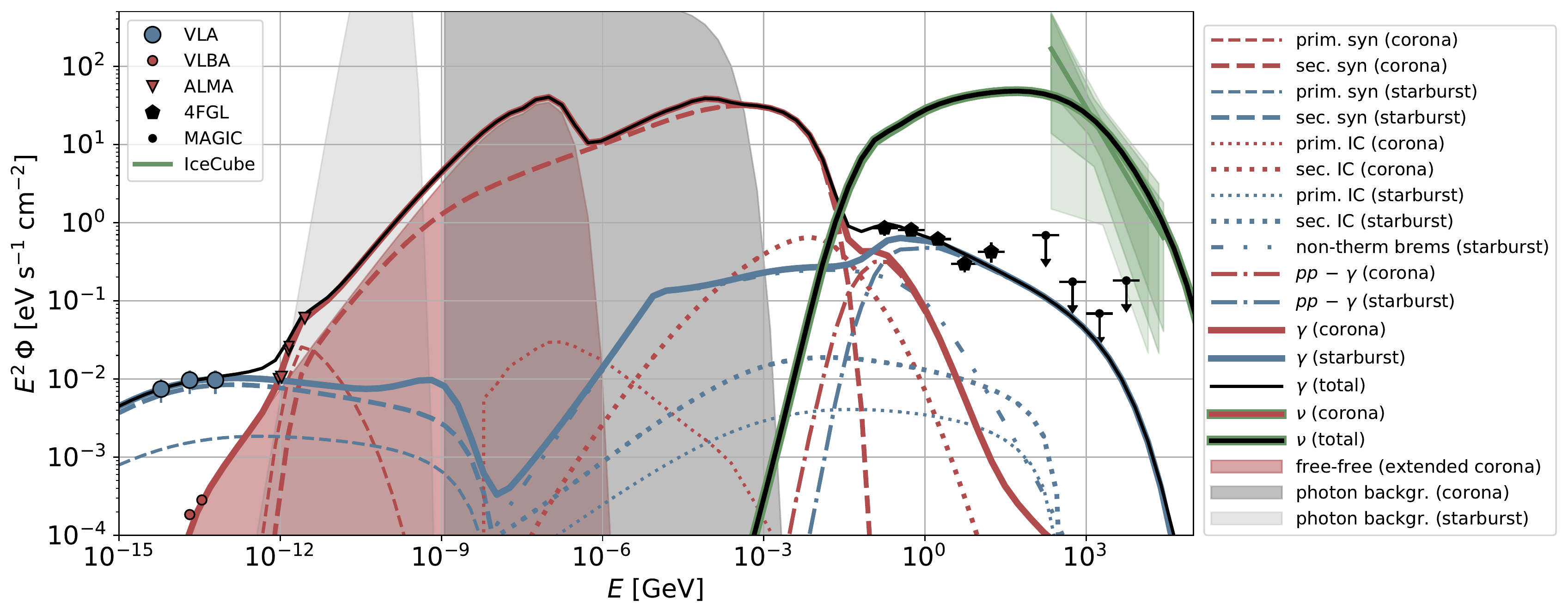}
    \caption{The model predictions of the \emph{photon and neutrino} SED of NGC\,1068 with respect to the data---red markers refer to a beam size of $\sim (0.02-0.06)\,\text{arcsec}$, and black or blue markers indicate a beam size of $\gtrsim 10\,\text{arcsec}$. The light red area shows the free-free emission from the extended corona. The dark grey area indicates the \emph{internal} flux of the background (target) photon fields (disk- and torus emission as well as Comptonized X-rays of the AGN corona) of the central AGN and the light grey area indicates the thermal IR emission by dust grains of the starburst region. Note that the torus attenuation is not taken into account here.}
    \label{fig:SED_details} 
\end{figure}

In the high (low) coronal X-ray luminosity case, the best-fit parameters of the coronal gas density, radius and magnetic field strength correspond to an optical Thomson depth of $\omega_{\rm T}\simeq 2.6\,(4.4)$ as well as a coronal plasma beta of $\beta\simeq 0.1$ (in both cases). Here, a CR luminosity of $L_{\rm CR}=9.55\,(2.07)\times 10^{43}\,\text{erg}\,\text{s}^{-1}$ is needed which is about $4.5\,\%$ ($2.1\,\%$) of the Eddington luminosity. Further, we obtain a coronal CR-to-thermal gas density pressure of $P_{\rm CR}/P_{\rm gas}=0.30\,(0.10)$. We checked that also for the case of Kraichnan turbulence ($\varkappa=3/2$) quite similar fit results can be obtained, however, some of the parameter values change---such as e.g.\ slightly larger plasma beta values (of up to 0.2) and smaller CR-to-thermal gas density pressure ratios (of 12 and 6\% in case of a high and low X-ray luminosity, respectively) for the corona. 
Fig.~\ref{fig:SED_details} shows that the minimal $\chi^2$ fit to the data yields an almost perfect agreement with the data---except for the 4FGL data point at the highest energy. 
The large scale radio data are described by synchrotron radiation of predominantly secondary electrons from the starburst region, whereas the small scale radio/IR data result from the coronal synchrotron radiation of mostly primary electrons. Due to the optical thickness by synchrotron self-absorption, this spectrum cuts off sharply towards small frequencies in the FIR. At the high energy end of the spectrum, where the impact of the torus attenuation vanishes, the $\gamma$-ray contribution of the corona and the starburst are at about the same level at about $50\,\text{MeV}$, so that at low $\gamma$-ray energies both regions are needed to describe the multi-wavelength data. At about few $\times 100\,\text{MeV}$ the coronal $\gamma$-ray emission becomes subdominant and the $\gamma$-rays of the starburst ring are sufficient to explain the observed $\gamma$-ray data above about $1\,\text{GeV}$. Its high $\gamma$-ray luminosity is mostly a consequence of the estimated SN rate of $0.5\,\text{yr}^{-1}$. However, we verified that even for a significantly lower rate the data can still be explained---although the chi-squared value increases due to increasing deviations in the radio and $\gamma$-ray band. In this case a higher coronal $\gamma$-ray flux is needed that compensates the lack of $\gamma$-rays from the starburst ring and explains the data up to about $1\,\text{GeV}$. 
The observed IceCube neutrinos can be explained  by the corona, as the neutrinos---in contrast to the associated $\gamma$-rays---are able to leave this central region. However, in the the best-fit case of a low coronal X-ray luminosity ($L_X=0.9\times 10^{43}\,\text{erg/s}$) the resulting neutrino flux at $\lesssim 1\,\text{TeV}$ is about a factor of five smaller than what is shown in Fig.~\ref{fig:SED_details} (but still matches the potential IceCube flux at $28\,\text{TeV}$). Independent of the adopted coronal X-ray field or the particular fit scenario, our model suggests a hardening of the neutrino flux below about $1\,\text{TeV}$. 

An alternative fit scenario with $\chi^2\sim 14$, that enables a higher neutrino flux at $\lesssim 1\,\text{TeV}$ for the low X-ray luminosity case is briefly summarized in the following: 
Using a higher injection fraction ($f_{\rm inj}\sim 0.06$) and a smaller radius ($\sim 95\,\mathcal{R}_{\rm s}$) for the corona, it becomes possible to match the potential neutrino flux also at around $1\,\text{TeV}$. However, in that case the corona yields a higher $\gamma$-ray flux at a few $\times 100\,\text{MeV}$, so that the starburst ring needs to be negligible at these energies. Hence a somewhat smaller gas density ($n_{\rm gas}\sim 100\,\text{cm}^{-3}$) and harder initial CR spectrum ($s\sim 2.1$) in the starburst ring is needed, but still the data at about $100\,\text{MeV}$ is slightly overshot due to the additional minor contribution by the starburst ring. At low energies the data is still explained quite accurately, in which the coronal IR emission results from synchrotron radiation of secondary electrons, whereas primary synchrotron emission is no longer present. However, this scenario yields a much higher CR pressure ($P_{\rm CR}/P_{\rm gas}\simeq0.50$) in the corona, so that altogether we consider this alternative scenario to be less likely than the best-fit scenarios that have been described previously. 

\section{Conclusions and discussion} \label{sec:conclusions}
In this work, we introduced a spatially homogeneous, spherically symmetric, steady state two-zone model for AGN-starburst composite galaxies. Using the multi-messenger data of NGC\,1068 from the radio up the the $\gamma$-ray band as well as its recent indications of high-energy neutrino emission, we present a first application of this model. Hereby, we perform a global parameter optimization within the $(2\times6)$-dimensional parameter space and manage to perfectly explain \emph{all} data---except for some minor deviations of the $\gamma$-ray flux at about $10\,\text{GeV}$. So, the $\gamma$-ray emission above a few $\times 100\,\text{MeV}$ results predominantly from the starburst region, whereas the high-energy neutrinos at TeV energies must originate from the coronal region. As already discussed in other works \citep[e.g.][]{a5:Murase+2020,Inoue+2020,Kheirandish+2021} the corona is optically thick for the associated $\gamma$-rays, which introduced a cascade of secondary electrons that dominate the emission at $0.1\,\text{eV}\lesssim E_\gamma\lesssim 100\,\text{MeV}$ via synchrotron and IC radiation---in addition to the strong free-free emission of the hot gas. In contrast to these other works we however manage to explain the high-energy neutrino emission by using an acceleration scenario where the CRs are scattered off stochastically by Alfv\'enic turbulence that shows either a small spectral index ($\varkappa\lesssim 3/2$) or a turbulence strength parameter $\eta\sim 1$. Hence, there is no need for an alternative acceleration process such as magnetic reconnection, as studied extensively in \cite{Kheirandish+2021}, even though, such an alternative acceleration scenario in the AGN corona would relax the need for strong Alfv\'enic turbulence. 
Further, the resulting gas density, radius and magnetic field strength of the corona yield a rather high optical Thomson depth of $\omega_{\rm T}\gtrsim 1$ as well as a low plasma beta of $\beta\sim 0.1$, which however, is within the range of expectations \citep[e.g.][]{Ricci+2018, MillerStone2000}. Some of our best-fit scenarios suggest a rather large CR pressure of about $30\%$ of the thermal gas pressure, hence, a huge amount of the gravitational binding energy goes into CRs. But this becomes less extreme if we account for the additional energy that is supplied by the disk. 

Finally, we manage to explain all data well in case of a strongly magnetized corona and a starburst ring with a high supernova rate ($\sim 0.5\,\text{yr}^{-1}$). 
Such a full multi-messenger fit from radio to TeV energies in photons plus the potential neutrino flux has not been attempted before. In particular, using the pure AGN core model has difficulties explaining the full high-energy signatures \citep[][]{a5:Murase+2020, Inoue+2020, Kheirandish+2021}. 
Including the additional contribution from the starburst ring obviously helps to explain the photon emission above about $100\,\text{MeV}$, but also with respect to the coronal high-energy neutrino emission the detailed fitting approach enables us to find a much better agreement to the potential neutrino flux. 

In total we showed that the broadband multi-messenger data of NGC\,1068 can only be explained if we account for the non-thermal emission by the outer starburst ring as well as the inner corona. 
However, we are not able to explain the VLBA radio data of the central region by the inner corona region, neither via free-free emission (due to the high electron temperature), nor via synchrotron radiation (due to the optically thickness at these frequencies for a magnetic field strength of $> 10\,\text{G}$). Therefore, we followed the common assumption \citep[see e.g.][]{Roy+1998,Gallimore+2004,Inoue+2020} and introduced the extended coronal region (extending up to $0.7\,\text{pc}$) to explain these data. Here we suppose that this extended corona is filled with a thermal gas with an electron temperature of $10^6\,\text{K}$ that emits free-free radiation and becomes optically thick at about $5\,\text{GHz}$, which yields an appropriate agreement with the VLBA data. Based on the effect of radiation pressure compression on an ionized gas \cite{BaskinLaor2021} recently showed that the brightness temperature of a dusty gas is limited to $2\times 10^5\,\text{K}$ at $5\,\text{GHz}$. In addition, they showed that a hot free-free emitting gas (with electron temperatures of $\gtrsim 10^7\,\text{K}$) also over produces the observed X-ray luminosity of NGC 1068. Therefore, they exclude optically thin free-free emission in NGC 1068, on the sub pc scales. We noticed, that due to the given maximal extension of this region an electron temperature of at least $10^6\,\text{K}$ is needed and the upper flux limit at $1.4\,\text{GHz}$ can only be satisfied if the free-free emission already becomes optically thick at about $5\,\text{GHz}$. In that case the free-free emitting gas is subdominant at X-rays and necessary electron temperature could be realized for a dustless gas even under consideration of radiation pressure compression.

As previous models, our model is limited with respect to the spatial description of the two emission zones, so that inhomogeneities and magnetic field structures cannot be taken into account. Hence, a more accurate treatment of the spatial structures as well as the three-dimensional transport might change some of the details of these results and should be taken into account in future work. 

In general, more data in particular in the range of about $(1-100)\,\text{MeV}$ would be very useful to further constrain the model. At lower energies the coronal emission is expected to be attenuated by the torus, so that it would become necessary to account for the physical processes in the torus region, if the data cannot resolve the sub-torus structures. As the coronal region is a perfect CR calorimeter, we do not expect any additional non-thermal emission from that region. Hence, it is not expected that the model prediction benefits from the inclusion of the torus, as this involves another significant expansion of the parameter space. But in case CR protons get also accelerated up to TeV energies in the torus---such as by winds from the coronal region that impact the torus and trigger shocks as proposed recently by \cite{InoueCerrutiMuraseLiu2022}---the observed GeV photons could also originate from hadronic pion production with the torus gas. 

Another sub-structure of NGC\,1068---which we do not take into account---is its jet that has been observed by centimeter radio observations on scales of a few $\times 100\,\text{pc}$ \citep[e.g.][]{WilsonUlvestad1982,Gallimore+2004,Gallimore+2006}. Therefore, we cannot exclude that there is some minor contamination of the considered large scale radio data by the jet. In addition, this jet has previously been discussed \citep{Lenain+2010,Lamastra+2019} as a possible origin of the $\gamma$-ray signal, as the necessary $\gamma$-ray luminosity of NGC 1068 is typically too high to be explained only by the \emph{nuclear} starburst activity \citep[see also][]{a5:eichmann2016,a5:Yoast-Hull-etal-2014}. Based on the bolometric IR luminosity of the starburst ring we estimate an average SN rate of $0.5\,\text{yr}^{-1}$, which is almost an order of magnitude higher than what has been supposed for the nuclear starburst activity and right within the range of $(0.1-1)\,\text{yr}^{-1}$ that has been suggested by \cite{Mannucci+2003,Wilson+1991}. Thus, if about $(4-5)\%$ of that SN energy is converted into CRs---which is similar to what has been found in numerical simulations \citep[e.g.][]{Haggerty+2020}---there is no need for the non-thermal jet emission to explain the $\gamma$-ray data. But since the jet as well as the torus region is not taken into account in this work, we cannot exclude that they provide some contribution to the observed $\gamma$-ray flux.

\begin{acknowledgments}
We would like to thank the anonymous referee for constructive comments that helped to improve the original version of this paper. Thanks also go to M.~Kachelrie\ss, M.~Zacharias, as well as E.~Kun for fruitful discussions in terms of different aspects of this work. Further, we acknowledge funding from the German Science Foundation DFG, within the Collaborative Research Center SFB 1491 ``Cosmic Interacting Matters - From Source to Signal". In addition, BE acknowledges support by the DFG grant EI~963/2-1.
\end{acknowledgments}

\vspace{5mm}

\software{Some of the results in this paper have been derived using the software packages Numpy \citep{vanDerWalt2011}, Scipy \citep{2020SciPy-NMeth}, Pandas \citep{mckinney-proc-scipy-2010}, Matplotlib \citep{Hunter:2007}, Seaborn \citep{Waskom2021}.
          }

\appendix

\section{Energy losses}\label{sec:energy_losses}
CR electrons lose energy due to bremsstrahlung, ionization, synchrotron radiation and inverse
Compton scattering, while for CR protons the processes to consider are synchrotron radiation, hadronic pion production, Bethe Heitler pair production and photopion processes. Before illustrating these energy loss rates, the Lorentz factor $\gamma$ of CR electrons and protons, respectively, and the dimensionless velocity $\beta$ are introduced: 
\begin{equation}
    \gamma \equiv \frac{T}{E_{0}}+1 \qquad \beta \equiv \sqrt{1-\frac{1}{\gamma^2}}
\end{equation}
Here $T$ stands for the relativistic kinetic energy of a CR particle with a rest energy $E_{0}$. In the following, we add an index to those quantities to specify that they refer to the CR electron (e) or protons (p).
\medskip

For a fully ionized medium with density $n_Z$, the relativistic electron bremsstrahlung $
e + p \longrightarrow e + p + \gamma$
energy loss rate is given by the expression \citep{Dermer+2009_book}
\begin{equation}
    \tau_{\rm brems}^{-1} = \frac{3}{2\pi}\alpha_Fc\sigma_T \left(\ln2\gamma_{\rm e}-\frac{1}{3}\right)\left(\sum_Z n_ZZ\left(Z+1\right)\right)\,,
\end{equation}
where $\alpha_F$ is the fine structure constant, $\sigma_T$ is the Thomson cross section and $Z$ is the atomic charge number of the particles characterising the medium.
\medskip

Another way high energy particles lose energy is by ionization and excitation to bound atomic levels of the matter they travel through. In a fully ionized plasma, the energy loss is due to scattering of individual plasma electrons and due to the excitations of large-scale systematic or collective motions of many plasma electrons. The energy loss rate due to this process is \citep{Schlickeiser2002_book}
\begin{equation}
  (\tau_{\rm C}^{\rm (e)})^{-1} = \frac{3}{4\gamma_{\rm e}}c\sigma_Tn_e\left(74.3+\ln\frac{\gamma_{\rm e}}{n_e}\right)\,,
\end{equation}
where $n_e$ denotes the thermal electron density, which is about equal to the gas density in a fully ionized plasma.
\medskip

The low energy part of the NGC 1068 emission spectrum is strongly affected by synchrotron radiation, created when charged particles move through a magnetic field. The electron synchrotron cooling rate is given by \citep{RevModPhys.42.237}
\begin{equation}\label{synch_e}
    \tau_{\rm syn}^{-1} = \frac{4\,c\sigma_T}{3E_{e,0}}\,\frac{B^2}{8\pi}\,\gamma_{\rm e}\beta_{\rm e}^2\,.
\end{equation}
The inverse Compton process involves the scattering
of low energy photons to high energies by ultra-relativistic electrons so that the
photons gain and the electrons lose energy. Integrating the inverse Compton power of photons 
over all scattered photon energies $E_\gamma$, we obtain the energy loss of a single relativistic electron due to inverse Compton scattering \citep{Schlickeiser2002_book} 
\begin{equation}
    \tau_{\rm ic}^{-1} = \frac{3\sigma_T c E_{e,0}\beta_{\rm e}^2\,}{4\,\gamma_{\rm e}} \int_0^\infty \diff E^{\prime}\frac{n(E^{\prime})}{E^{\prime}}\int_0^1 \diff q \frac{\Gamma^2 q}{(1+\Gamma q)^3}F_{\rm KN}(q, \Gamma)
\end{equation}
where \citep{RevModPhys.42.237} 
\be
F_{\rm KN}\left(q, \Gamma\right)=2q\ln(q) +1+q-2q^2+\frac{\left(\Gamma q\right)^2\left(1-q\right)}{2\left(1+\Gamma q\right)}
\ee
with $\Gamma=4E^{\prime} \gamma_{\rm e}/(m_e c^2)$ and $q=E/[\Gamma \left(\gamma_{\rm e} m_e c^2-E\right)]$. The terms $E^{\prime}$ and $E$ are, respectively, the energies of the photon before and after the inverse Compton scattering.
\medskip

When considering protons, the energy loss rate due to synchrotron radiation needs to be rescaled due to the decrease of the cross-section, so that
\begin{equation}\label{synch_p}
    \left(\tau_{\rm syn}^{(p)}\right)^{-1} = \left(\frac{m_e}{m_p}\right)^3 \frac{4\,c\sigma_T}{3E_{p,0}}\,\frac{B^2}{8\pi}\,\gamma_{\rm p}\beta_{\rm p}^2\,.
\end{equation}
Also the energy loss rate of a fast proton due to Coulomb interactions with the fully ionized plasma has a different cross-section yielding \citep{Schlickeiser2002_book}
\begin{equation}
  (\tau_{\rm C}^{\rm (p)})^{-1} \simeq 3.1\times 10^{-7}\,\left(\frac{n_e}{1\,\text{cm}^{-3}}\right)\,\left(\frac{T_p}{1\,\text{eV}}\right)^{-1}\,\frac{\beta_p^2}{x_{\rm m}^3+\beta_p^3}\,\, \text{s$^{-1}$}\,,
\end{equation}
where $x_{\rm m}=0.0286\,(\theta_e/2\times 10^6\,\text{K})^{1/2}$. 
In addition, relativistic protons can interact with the gas protons and produce pions, in the so called hadronic pion production process: $ p + p \longrightarrow \pi + X$ ($X$ is anything else created in the $pp$ collision \citep[e.g.][]{Becker:2007sv,Dermer+2009_book}). The particles energy loss rate can be approximated in the range $1.2\,\text{GeV}< E \leq 10^8\,\text{GeV}$ by \citep{Krakau+2015}
\begin{equation}
    \tau_{\rm pp}^{-1} = 4.4\cdot 10^{-16}\,H(\gamma_{\rm p}-1.3)\,\left(\frac{n_{\rm gas}}{1\,\mathrm{cm}^{-3}}\right) \gamma_{\rm p}^{0.28}\,\beta_{\rm p}^{0.56}\,\cdot(\gamma_{\rm p}+187.6)^{-0.2} \,\, \text{s$^{-1}$}
\end{equation}
where $n_{\rm gas}$ is the interstellar gas density and $H()$ denotes the Heaviside function to account for these losses only above $1.2\,\text{GeV}$. Note that we included the $\beta_{\rm p}$ dependence to enable an extrapolation towards mildly relativistic energies.
\medskip

Proton interactions with the background (target) photons can produce $e^{\pm}$ pairs ($p + \gamma \longrightarrow p + e^+ + e^-$), which can lead to electromagnetic cascades. This
process is called Bethe-Heitler pair production and it is described by the characteristic particles energy loss rate of \citep{zheng2016bethe}
\begin{equation}
  \tau_{\rm BH}^{-1} = \frac{m_p^2m_e^2 c^9}{2E_p^2}  \int_{E_{\gamma,{\rm min}}}^{\infty}\frac{n_{\gamma}(E_{\gamma})}{E_{\gamma}^2}\diff E_{\gamma}\int_{E_{\rm min}^{\prime}}^{E_{\rm max}^{\prime}} \sigma_{p\gamma,e}\left(E^{\prime}\right)E^{\prime}\diff E^{\prime}\,, 
  \label{eq:BHlossRate}
\end{equation}
\noindent where $E^{\prime} = \gamma_{\rm p} E\left(1-\beta_{\rm p}\cos\lambda\right)$ is the energy of the photon in the rest frame of the proton with the angle between the proton and photon directions $\lambda$ and the proton velocity $\beta_{\rm p}$ is in units of $c$. The terms $E_{\rm min}^{\prime}$ and $E_{\rm max}^{\prime}$ correspond respectively to 1 MeV/$m_ec^2$ and $2\gamma_{\rm p}E$, while $E_{\gamma,{\rm min}}$ = 1 MeV $/(2\gamma_{\rm p})$.
\medskip

When the interaction between a proton and a target photon produces a pion ($p + \gamma \longrightarrow \pi$ + p), the typical energy loss rate of the initial particle is given by \citep{Dermer+2009_book}
\begin{equation}
    \left(\tau_{p\gamma}^{\pi}\right)^{-1} \simeq \frac{c}{2\gamma_{\rm p}^2} \int_0^{\infty}\frac{n_{\gamma}\left(E_{\gamma}\right)}{E_{\gamma}^2}\diff E_{\gamma}\int_0^{2\gamma_pE} E^{\prime}\sigma_{p\gamma,\pi}\left(E^{\prime}\right)K_{p\gamma}\left(E^{\prime}\right)\diff E^{\prime}
\end{equation}

\noindent where $\gamma_{\rm p}$ is the proton Lorentz factor and $K_{p\gamma}(E^{\prime})$ is the fraction of energy lost by the ultrarelativistic proton ($\gamma_{\rm p} \gg 1$ and $\beta_{\rm p} \longrightarrow 1$) in the interaction, therefore the inelasticity of the collision.

\section{Emissivities}\label{sec:emissivities}
The emissivities (in units of $\text{cm$^{-3}$ s$^{-1}$ eV$^{-1}$}$) of secondary particles (photons, electrons/positrons and neutrinos) with an energy $E$ that we introduce in the following account for (i) synchrotron radiation, inverse Compton scattering and bremsstrahlung of CR electrons; (ii) hadronic and photo-hadronic pion production as well as Bethe-Heitler pair production of CR protons; (iii) free-free emission of the thermal gas; and (iv) $\gamma\gamma$ pair production.
\medskip

The synchrotron emission by CR protons is negligible (see Equation (\ref{synch_e}) and (\ref{synch_p})), so that only synchrotron radiation by CR electrons is considered in the following. Its spectral power is given by \cite[e.g.][]{RevModPhys.42.237}
\begin{equation}
    P_{\rm syn}(\nu,\gamma_{\rm e}) = P_0\left(\frac{\nu}{\nu_s\gamma_{\rm e}^2}\right)^\frac{1}{3}\exp\left({\frac{\nu}{\nu_s\gamma_{\rm e}^2}}\right)
    \label{eq:synPower}
\end{equation}
with $P_0$ = 2.65$\times$10$^-$$^1$$^0$ $\left(B/1 \text{G}\right)$ eV s$^{-1}$Hz$^{-1}$ and $\nu_s$ = 4.2$\times$10$^6$ $\left(B/1 \text{G}\right)$ Hz. The isotropic spontaneous synchrotron emission coefficient of the relativistic electron distribution yields
\begin{equation}\begin{split}
    \epsilon_{\rm syn}\left(\nu\right) =  &\int_1^\infty \diff \gamma_{\rm e}\,\, n_e\left(\gamma_{\rm e}\right)P_{\rm syn}\left(\nu, \gamma_{\rm e}\right) = P_0\left(\frac{\nu}{\nu_s}\right)^{\frac{1}{3}}\cdot\int_1^\infty \diff\gamma_{\rm e}\,\, \gamma_{\rm e}^{-\frac{2}{3}}\exp\left(-\frac{\nu}{\nu_s\gamma_{\rm e}^2}\right)n_e\left(\gamma_{\rm e} \right)\,,
    \end{split}
\end{equation}
where $n_e(\gamma_{\rm e})$ is the differential CR electron density.
\medskip

The emissivity function for bremsstrahlung radiation produced by CR electrons is  \citep{stecker1971cosmic}
\begin{equation}
    \epsilon_{\rm brems} \left(E\right) = \frac{c n_{\rm gas}\sigma_{\rm brems}}{E}\int_{E/(m_ec^2)}^\infty \diff\gamma_{\rm e}\, n_e\left(\gamma_{\rm e}\right)\,, 
\end{equation}
where $\sigma_{\rm brems} = 3.38\times10^{-26}$ cm$^2$.
\medskip

Charged pions formed by the photomeson process decay into leptons and neutrinos, and neutral pions decay into $\gamma$-rays. The emissivity of these secondaries, assuming isotropy of the photon and ultra high energetic cosmic ray proton spectra, is described by the following expression \citep{Dermer+2009_book}:
\begin{equation}
    \epsilon_{\pi\gamma}^{i}\left(E\right) = \frac{\zeta_i c\sigma_{\pi\gamma}^{i} n_p\left(\bar{\gamma}_p\right)}{16\pi\chi_i m_p c^2\bar{\gamma}_p^2}\int_{\frac{E_{l}^{\prime}}{2\bar{\gamma}_p}}^{\infty} \diff E_{\gamma}\frac{n_{ph}(E_{\gamma})}{E_{\gamma}^2}\left\{\big[\min\left(2\bar{\gamma}_p E_{\gamma}, E_u^{\prime}\right)\big]^2-E_{l}^{\prime 2}\right\}\,.
\end{equation}
\begin{table}
  \caption{Multiplicities $\zeta$ and mean fractional energies $\chi$ of secondaries formed in photomeson production.}
  \centering 
  \begin{threeparttable}
    \begin{tabular}{ccc}
    Species  & Single $\pi$ & Multi $\pi$\\
     \midrule
Neutrinos  &   
$\zeta_\nu^s=3/2 \quad \chi_\nu^s=0.05$        &   $\zeta_\nu^m=6 \quad \chi_\nu^m=0.05$                       \\
    Electrons & $\zeta_e^s=1/2 \quad \chi_e^s=0.05$ & $\zeta_e^m=2 \quad \chi_e^m=0.05$ \\ 
    $\gamma$-rays & $\zeta_\gamma^s=1 \quad \chi_\gamma^s=0.1$ & $\zeta_\gamma^m=2 \quad \chi_\gamma^m=0.1$ \\ 
    \midrule
    \end{tabular}
\end{threeparttable}
\label{tab:zeta_chi}
\end{table}
Hereby, for secondaries generated by single pion production we use $\sigma^s_{\pi\gamma}$ = 340 $\mu$b, $\tilde{E}_l^{\prime}$ = 390, $\tilde{E}_u^{\prime}$ = 980, whereas for those produced in multi pion production we adopt $\sigma^m_{\pi\gamma}$ = 120 $\mu$b, $\tilde{E}_l^{\prime}$ = 980, $\tilde{E}_u^{\prime} \longrightarrow \infty$. Further, $\zeta_i$ denotes the multiplicity of secondary $i$, $\chi_i$ is the mean fractional energy of the produced secondary compared to the incident primary proton (see Table \ref{tab:zeta_chi}), $n_p(\bar{\gamma}_p)$ is the differential intensity of CR protons with a Lorentz factor $\bar{\gamma}_p \equiv E/(\chi_im_pc^2)$. The dimensionless target photon energy $E_{\gamma}$ is described in units of $E_{e,0}$.
\medskip

The free-free radiation emission of a thermal gas with about the same number density of electrons and ions is defined as \citep{padmanabhan2000theoretical}
\begin{equation}
    \epsilon_{ff} \left(E\right) = \frac{32\pi}{3}\sqrt{\frac{2\pi}{3}}\,n_{\rm gas}^2\,\frac{q_e^6}{m_ec^2}\sqrt{\frac{1}{k_B\theta m_ec^2}}\exp\left(-\frac{E}{k_B\theta}\right)g_{ff}\,,
    \label{eq:freefreeEmission}
\end{equation}
where $\theta$ is the temperature and $g_{ff}$ is the so called Gaunt factor, which is given by  \citep{padmanabhan2000theoretical}:
\begin{equation}
    g_{ff} = 
    \begin{cases}
    \frac{\sqrt{3}}{\pi}\ln\left(\frac{2k_{B}\theta}{\left(1.78 m_{e}\right)^{3/2}}\,\frac{2m_{e}}{1.78q_{e}^2\pi E/h}\right)\,,\quad&\text{for }\theta<8.9\times10^5\,\text{K}\,,\\
    \frac{\sqrt{3}}{\pi} \ln\left(\frac{4k_{B} \theta}{1.78E}\right)\,,\quad&\text{for }\theta\geq 8.9\times10^5\,\text{K}\,.
    \end{cases}
\end{equation}
For the hadronic pion production process the emissivity of secondary particles of type $a$---either photons ($\gamma$), neutrinos ($\nu$) or electrons ($e$)---is given by \citep{koldobskiy2021energy}
\begin{equation}
    \epsilon_{ pp}^a\left(E\right) = c\, n_{gas}\int_E^{\infty} \diff E^{\prime}\frac{\diff\sigma_a}{\diff E}\left(E^{\prime},E\right)n_p\left(E^{\prime}\right)\,,
\end{equation}
in the case of an homogeneous source, where $n_{gas}$ denotes the number density of target protons. The differential cross-section $\diff\sigma_a/\diff E$ is adopted from the Kamae et al.\ parametrisation \citep{kamae2006parameterization} below a threshold energy of 4 GeV, and above that energy by the \texttt{AAfrag} parametrisation \citep{koldobskiy2021energy}. 
\medskip

For inverse Compton radiation the $\gamma$-ray emissivity is determined by \citep{Schlickeiser2002_book}
\begin{equation}
    \epsilon_{\rm ic}\left(E
    \right) = \frac{3c\sigma_T}{4}\int_0^\infty \diff E^{\prime} \frac{v\left(E^{\prime}\right)}{E^{\prime}}\int_{\gamma_{\rm min}}^\infty \diff\gamma_{\rm e} \frac{n_e\left(\gamma_{\rm e}\right)}{\gamma_{\rm e}^2}F_\text{KN}\left(q,\Gamma\right)\,,
\end{equation}
where 
$$\gamma_{\rm min}= \frac{E}{2m_ec^2}\left[1+\sqrt{1+\frac{m_e^2c^4}{E^{\prime}E}}\right]$$ and $F_\text{KN}(q,\Gamma)$ is previously defined in the inverse Compton energy loss rate.\medskip

The Bethe-Heitler process emissivity is given by \citep{zheng2016bethe}
\begin{equation}
    \epsilon_{\rm BH}\left(E\right) = 2\frac{m_p}{m_e}n_p\left(\frac{m_p}{m_e}E\right)\left[\tau_{\rm BH}\left(\frac{m_p}{m_e}E\right)\right]^{-1} 
\end{equation}
where $n_p$ is the differential CR proton number density and $\tau_{\rm BH}^{-1}$ is the energy loss rate as introduced in Eq.~(\ref{eq:BHlossRate}). 
\medskip

Another important process to take into account in terms of the production of secondary electrons and positrons is $\gamma\gamma$ pair production ($\gamma + \gamma \longrightarrow e^+ + e^-$) as the corona becomes optically thick at high energy. Here, $\gamma$-rays that are in our case predominantly generated by the CR protons via hadronic pion production (with an emissivity $\epsilon_{\rm pp}^{\gamma}$) interact with the thermal background photons (which are Comptonized up to X-ray energies within the corona) and introduce another source of electrons/ positrons. 
According to \cite{bottcher2013leptonic} the steady state production rate of these $e^{+}/e^{-}$ pairs can be estimated by 
\begin{equation}
\epsilon_{\gamma\gamma}\left(\gamma_{\rm e}\right) = f_{abs}\left(\tilde{E}_1\right)\cdot\epsilon_{\rm pp}^{\gamma}\left(\tilde{E}_1\right) + f_{abs}\left(\tilde{E}_2\right)\cdot\epsilon_{\rm pp}^{\gamma}\left(\tilde{E}_2\right) 
\end{equation}
where $\tilde{E}_1$ = $\gamma_{\rm e}/f_{\gamma}$ and $\tilde{E}_2$ = $\gamma_{\rm e}/(1-f_{\gamma})$ denote the characteristic dimensionless $\gamma$-ray energies (in units of $m_ec^2$), and the absorption fraction $f_{abs} \equiv 1 - [{1-\exp(-\omega_{\gamma\gamma}\left(E\right))}]/{\omega_{\gamma\gamma}\left(E\right)}$ is determined from the given optical depth $\omega_{\gamma\gamma}\left(E\right)$ due to this process (see Eq.~\ref{eq:gammgammaAbs}). We assume that one of the produced particles obtains the major fraction $f_{\gamma}$ of the photon energy, where $f_{\gamma}$ = 0.9 is found from Monte Carlo simulations. Hence, an electron/positron pair is produced with energies $\gamma_1$ = $f_{\gamma}E$ and $\gamma_2$ = $(1-f_{\gamma})E$. 

\section{Absorption coefficients}\label{sec:absorption_coeff}
The effective optical thickness for an homogeneous medium is given by $\omega=R\,(\alpha_{\rm syn}+\alpha_{\rm ff}+\alpha_{\gamma\gamma})$, where, for the considered astrophysical environments, we need to account for the synchrotron-self absorption ($\alpha_{\rm syn}$), the free free absorption ($\alpha_{\rm ff}$) and the $\gamma\gamma$ pair production ($\alpha_{\gamma\gamma}$) process.
\medskip

The synchrotron self-absorption can be described by \citep{Schlickeiser2002_book}
\begin{equation}
    \alpha_{\rm syn}(\nu,\theta) = -\frac{c^2}{8\pi m_e c^2\nu^2}\int_1^{\infty}\diff \gamma_{\rm e}\,\,\gamma_{\rm e}^2P_{\rm syn}\left(\nu,\gamma_{\rm e}\right)\frac{\diff}{\diff\gamma_{\rm e}}\left[\gamma_{\rm e}^{-2}n_{\rm e}(\gamma_{\rm e})\right] 
\end{equation}
where $n_{\rm e}(\gamma_{\rm e})$ denotes the differential number density of CR electrons, which are assumed to be isotropic, and $P_{\rm syn}\left(\nu,\gamma_{\rm e}\right)$ is the total spontaneously emitted spectral synchrotron power of a single relativistic electron that has already been introduced in Eq.~(\ref{eq:synPower}).
\medskip

The thermal bremsstrahlung (free free) radiation emitted by an electron moving in the field of an ion---such as given by Eq.~(\ref{eq:freefreeEmission})---can subsequently also get absorbed by the associated absorption process. This so-called free free absorption coefficient is given by \citep{rybicki1991radiative}
\begin{equation}
    \alpha_{ff}\left(E\right) = \frac{4q_e^6}{3m_ehc}\left(\frac{2\pi}{3k_Bm_e}\right)^{0.5}\theta^{-0.5}n_{\rm gas}^2\cdot\left(1-\exp\Big(-\frac{E}{k_B\theta}\Big)\right)\frac{1}{\nu^3}g_{ff} 
\end{equation}
where $\theta$ and $n_{\rm gas}$ stand for the gas temperature and density, respectively. We assume a quasi-neutral gas with about the same number of ions and electrons.  
\medskip

In contrast to the previous processes the $\gamma\gamma$ pair production process attenuates the photon intensity typically at $\gamma$-ray energies. The corresponding absorption coefficient for the $\gamma\gamma$ pair can be approximated by \citep{Dermer+2009_book}
\begin{equation}
    \alpha_{\gamma\gamma}\left(E\right) = \frac{\pi r_e^2}{\tilde{E}_1^2}\int_{1/\tilde{E}_1}^{\infty}\diff\tilde{E} E^{\prime -2}n_{\gamma}(\tilde{E})\bar{\varphi}(s_{\rm cm}^{0})
    \label{eq:gammgammaAbs}
\end{equation}
where $r_e$ is the classical electron radius, $\tilde{E}_1=E/(m_ec^2)$ denotes the dimensionless photon energy passing through a background of photons with energy $\tilde{E}$, $n_{\gamma}\left(\tilde{E}\right)$ is the isotropic photon field and $\bar{\varphi}\left(s_{\rm cm}^{0}\right)$ is defined as
\begin{equation}
\bar{\varphi}(s_{\rm cm}^{0}) = 2\int_1^{s_{\rm cm}^{0}}\diff s_{\rm cm}\frac{s_{\rm cm}\sigma_{\gamma\gamma}(s_{\rm cm})}{\pi r_e^2}
\end{equation}
Here $s_{\rm cm}^{0} \equiv \tilde{E}\tilde{E}_1$ and the $\gamma\gamma$ pair production cross section $\sigma_{\gamma\gamma}$ dependent on the dimensionless interaction energy $s_{\rm cm}$ is given by
\begin{equation}
    \sigma_{\gamma\gamma}(s_{\rm cm}) = \frac{1}{2}\pi r_e^2\left(1-\beta_{\rm cm}^2\right)\big[\left(3-\beta_{\rm cm}^4\right)\ln\Big(\frac{1+\beta_{\rm cm}}{1-\beta_{\rm cm}}\Big)-2\beta_{\rm cm}\left(2-\beta_{\rm cm}^2\right)\big]
\end{equation}
where $\beta_{\rm cm} = (1-\gamma_{\rm cm}^{-2})^{1/2} = \sqrt{1-s_{\rm cm}^{-1}}$ and $\gamma_{\rm cm}=\sqrt{s_{\rm cm}}$ denotes the center-of-momentum frame Lorentz factor of the produced electron/ positron.

\bibliography{references}{}
\bibliographystyle{aasjournal}

\end{document}